RICE UNIVERSITY

# Sampling Techniques for Boolean Satisfiability

by

**Kuldeep Singh Meel**

A THESIS SUBMITTED
IN PARTIAL FULFILLMENT OF THE
REQUIREMENTS FOR THE DEGREE

**Master of Science**

APPROVED, THESIS COMMITTEE:

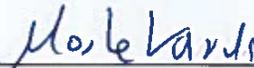

Moshe Y. Vardi, Chair
Karen Ostrum George Distinguished
Service Professor in Computational
Engineering, Rice University

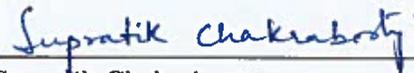

Supratik Chakraborty
Professor of Computer Science and
Engineering, IIT Bombay

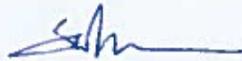

Swarat Chaudhuri
Assistant Professor of Computer Science,
Rice University

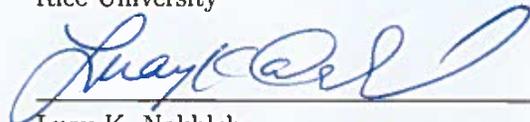

Luay K. Nakhleh
Associate Professor of Computer Science
and Biochemistry and Cell Biology, Rice
University

Houston, Texas

April, 2014

ABSTRACT

Sampling Techniques for Boolean Satisfiability

by

Kuldeep Singh Meel

Boolean satisfiability (SAT) has played a key role in diverse areas spanning testing, formal verification, planning, optimization, inferencing and the like. Apart from the classical problem of checking boolean satisfiability, the problems of generating satisfying uniformly at random, and of counting the total number of satisfying assignments have also attracted significant theoretical and practical interest over the years. Prior work offered heuristic approaches with very weak or no guarantee of performance, and theoretical approaches with proven guarantees, but poor performance in practice.

We propose a novel approach based on limited-independence hashing that allows us to design algorithms for both problems, with strong theoretical guarantees and scalability extending to thousands of variables. Based on this approach, we present two practical algorithms, UniWit: a near uniform generator and ApproxMC: the first scalable approximate model counter, along with reference implementations. Our algorithms work by issuing polynomial calls to SAT solver. We demonstrate scalability of our algorithms over a large set of benchmarks arising from different application domains.

Dedicated to my parents: Bhanwari and Ranjeet,

Madan Kajla and late Mahal Singh

# Acknowledgements

First and foremost, I owe my deepest gratitude to my advisors, Prof. Moshe Vardi and Prof. Supratik Chakraborty, for their steadfast support, guidance, encouragement and freedom to pursue my ideas. I would not have pursued graduate school had I not met Moshe in Mumbai in Dec'11. In every meeting with Moshe, I am truly inspired by his grand vision of research (not just limited to computer science) and his approach to scientific research solely motivated by intellectual curiosity. Supratik's meticulous nature along with his devotion to research has motivated me to pursue research with more passion than I could have imagined earlier. The following *shloka* from *Vishwasara Tantra* captures my experience with Moshe and Supratik better than I would be able to put in words.

अज्ञानतिमिरान्धस्य ज्ञानाञ्जनशालाकया ।
चक्षुरुन्मीलितं येन तस्मै श्रीगुरवे नमः ॥

*Salutations to the Guru who with corrylium stick of knowledge has opened the eyes of one blinded by disease of ignorance* (translation from [1])

I thank Prof. Swarat Chaudhuri for his sharp observations and constructive critique that improved this thesis significantly. I am grateful to Prof. Luay Nakhleh for generously agreeing to be on the thesis committee.

This thesis would not have a detailed empirical evaluation without a fine tuned version of CryptoMiniSAT thanks to Dr. Mate Soos and without Ajith John's help in benchmark generation and support of experimental infrastructure.

My technical writing, presentation skills and this thesis have tremendously benefited from my interactions with Dr. Tracy Volz and Dr. Jan Hewitt. I wish to thank Amit for his comments on early drafts of this thesis.

vI have been fortunate to share offices, hallways, lunches and interesting conversations over coffee and beer with wonderful colleagues over the last two years: Abhishek, Deepak, Dror, Giuseppe, Hrishikesh, Jianwen, Karthik, Milind, Morteza, Rahul, Ryan, Sailesh, Sonali, Sriraj, and Sumit. Given my dry humor , it must not have been easy for them to put up with me - especially when I strolled to their desks with the sole intention to crack some jokes. And out of the odd occasional research discussions, exciting collaborations were forged with Deepak, Dror, Karthik and Milind.

My social life outside Duncan Hall revolved around Mehul, Rakesh, Neeraj, Vaideesh, Rahul Rekhi and Amit. To Mehul and Rakesh, I can only say "Thank You" for lack of better words for your friendship, all the dinners (even the ones at Bawarchi and Ruggels Green), random conversations about everything in the universe and of course, beers. Transition to graduate school could not have been smoother, thanks to Vaideesh, Neeraj and Rahul Rekhi. I am grateful for counsel and friendship of Amit, who also continues to be my source for authentic Indian food. I would like to thank Devendra, Rho and Keliang for putting up with me as housemate.

I owe special acknowledgments to my friends from undergrad days on whom I continue to rely on in moments of self-doubt: Ashish, Devendra, Govind, Harshit, Pranav, Prashant, Ravi and Saurabh. I am grateful to Rahul, Sandeep and Vikash for being friends since the times I could count only up to 100.

I have been fortunate to have amazing teachers and mentors at different points in my life. I am forever beholden to my elementary school teachers, Madan Kajla and late Mahal Singh, for drilling in love for inquiry and reason at a young age. I am grateful to Prof. Rajkishore Nath, Prof. Peter Varman and Prof. Lin Zhong for encouragement, advice and writing several recommendation letters (including Rice PhD application) during undergraduate years.

I am ever indebted to my brother (Viru), uncles (Ramshwaroop, Ashok,Manoj and late Sripal Jakhar), aunts (Rajni and Pratibha) and grandparents (Rameshwar,

Mohini, Ramchandra and Banarasi), whom I have always taken for granted and whose abiding faith, love, encouragement and support has never abandoned my side.

Finally and most importantly, I would not have made it so far without uncountable selfless sacrifices of my parents, Bhanwari and Ranjeet. Even the times of financial hardship could not sway them from their sole focus to provide me with the best possible learning resources. I owe more to their constant love, support, encouragement and daily phone calls than words can possibly describe. It is to them and my elementary school teachers that I dedicate this thesis.

# Contents

# Illustrations

# List of Algorithms

# Chapter 1

# Introduction

Boolean satisfiability, also known as SAT, concerns determining the satisfiability of a given propositional formula. Since Cook showed SAT to be $\mathcal{NP}$-complete in 1971 [2] and Karp demonstrated polynomial-time reductions of several important problems to SAT [3], there has been strong theoretical and practical interest in the SAT problem. Specifically, SAT has played a key role in diverse areas spanning testing, formal verification, planning, optimization, inferencing, combinatorics and the like [4]. Apart from the classical problem of checking Boolean satisfiability, the problems of generating satisfying assignments uniformly at random, and of counting the total number of satisfying assignments have also attracted significant theoretical and practical interest over the years [5, 6, 7, 8, 9, 10, 11, 12, 13, 14]. This thesis focuses on the latter two problems: uniform generation and counting, and proposes new practical algorithms for solving them. Our first algorithm generates satisfying assignments of a propositional near-uniformly (explained in detail in Chapter 3). The core idea of this algorithm is then extended to approximately count the total number of satisfying assignments of a propositional formula. Unlike prior work, our algorithms provide strong theoretical guarantees and also scale to practical problem sizes. In the remainder of this Chapter, we briefly review motivating factors and previous work related to uniform generation and counting to put our contributions in context.

## 1.1 Uniform Generation

In this section, we motivate the problem of uniform generation of satisfying assignment. Functional verification constitutes one of the most challenging tasks in the development of modern hardware systems. Despite significant advances in formal verification over the last few decades, there is a huge mismatch between the sizes of industrial systems and the capabilities of state-of-the-art formal-verification tools [15]. Simulation-based verification techniques therefore dominate the functional-verification landscape [16]. A dominant paradigm in simulation-based verification is directed random testing, where an operational model of the system is simulated with a set of random test stimuli satisfying a set of *constraints* [9, 10, 11]. The simulated behavior is then compared with the expected behavior, and any mismatch is flagged as indicative of a bug. The constraints that stimuli must satisfy typically arise from various sources such as domain and application-specific knowledge, architectural and environmental requirements, specifications of corner-case scenarios, and the like. Test requirements from these varied sources are compiled into a set of constraints and fed to a constraint solver to obtain test stimuli. Developing constraint solvers (and test generators) that can reason about large sets of constraints is therefore acknowledged to be an important activity for industrial test and verification applications [17].

Despite the diligence and insights that go into developing constraint sets for generating directed random tests, the complexity of modern hardware systems makes it hard to predict the effectiveness of any specific test stimulus. It is therefore common practice to generate a large number of stimuli satisfying a given set of constraints. Since every stimulus is *a priori* as likely to expose a bug as any other stimulus, it is desirable to sample the solution space of the constraints uniformly or near-uniformly (defined formally in Chapter 3) at random [10]. A naive way to accomplish this is to first generate all possible solutions, and then sample them uniformly. Unfortunately, generating all solutions is computationally prohibitive and often infeasible in practical settings of directed random testing. For example, we have encountered systems of constraints where the expected number of solutions is of the order of $2^{100}$, and there is no simple way of deriving one solution from another. It is therefore interesting to ask: *Given a set of constraints, can we sample the solution space uniformly or near-uniformly, while scaling to problem sizes typical of testing/verification scenarios?* An affirmative answer to this question has implications not only for directed random testing, but also for other applications like probabilistic reasoning, approximate model counting and Markov logic networks [12, 13]. In practical applications, the requirement of strict uniformity can often be relaxed to some extent without affecting the quality of results. The relaxation in the requirement of uniformity is important, since known techniques for generating satisfying assignments with guarantees of strict uniformity such as [18], do not scale well in practice. Relaxed notions of uniformity, like "almost-uniform" or "near-uniform" are therefore important in practical applications of uniform generation. We discuss these relaxations of uniformity in Chapter 3. In this thesis, we discuss algorithms and tools that provide theoretical gaurantees of uniformity that conform to relaxed notions of uniformity and scale well in practice. Our another contribution is scalable algorithms and tools for a related problem: counting the satisfying assignments of given propositional formula, also known as model counting. We briefly review the motivating factors behind model counting in the following section.

## 1.2 Model Counting

Propositional model counting, also known as #SAT, concerns counting the number of models (satisfying truth assignments) of a given propositional formula. This problem has been the subject of extensive theoretical investigation since its introduction by Valiant [5] in 1979. Several interesting applications of #SAT have been studied in the context of probabilistic reasoning, planning, combinatorial design and other related fields [12, 19, 13]. In particular, probabilistic reasoning and inferencing have attracted considerable interest in recent years [20], and stand to benefit significantly from efficient propositional model counters. For example, the belief in a statement *s* for a knowledge base *B* with no explicit probabilistic information can be estimated by $\frac{\mathsf{M}(s \wedge B)}{\mathsf{M}(B)}$, where *s* and *B* are both encoded as propositional formulas and $\mathsf{M}(\cdot)$ is a function that gives the model count for an input formula.

Theoretical investigations of #SAT have led to the discovery of deep connections in complexity theory [6, 7, 8]: #SAT is #P-complete, where #P is the set of counting problems associated with decision problems in the complexity class $\mathcal{NP}$. Furthermore, $\mathsf{P}^{\mathsf{\#SAT}}$, that is, a polynomial-time machine with a #SAT oracle, can solve all problems in the entire polynomial hierarchy. In fact, the polynomial-time machine needs to make only one #SAT query to solve any problem in the polynomial hierarchy. This is strong evidence for the hardness of #SAT, which has been observed in the practice as well [20]. The techniques proposed for #SAT have been successfully used in small- to medium-sized problems, but scaling to larger problem instances has posed significant challenges in practice. Consequently, a large class of practical applications such as logistics,planning, inference has remained beyond the reach of exact model counters. This prompts us to explore the hardness of the relaxations of the exact counting (#SAT) from practical as well as theoretical perspective.

In many applications of model counting, such as in probabilistic reasoning, the exact model count may not be critically important, and approximate counts are sufficient. Even when exact model counts are important, the inherent complexity of the problem may force one to work with approximate counters in practice. Karp and Luby presented a fully polynomial randomized approximation scheme for counting models of a DNF formula [21]. While the DNF representation suits some applications, most modern applications of model counting (e.g. probabilistic inference) use the CNF representation [22]. Although *exact* counting for DNF and CNF formulae are polynomially inter-reducible, there is no known polynomial reduction for the corresponding *approximate* counting problems. In fact, Karp and Luby remark in [21] that it is highly unlikely that their randomized approximate algorithm for DNF formulae can be adapted to work for CNF formulae. Thus, there has been no prior implementation of approximate counters for CNF formulae *that scales in practice*. In this thesis, we provide algorithms and tools for approximate model counting that scale well in practice. Before we provide an overview of the contributions of this thesis, we explore the deep relationship between approximate model counting and uniform generation below.

## 1.3 Inter-reducibility of Approximate Counting and Almost Uniform Generation

The motivating factors for (almost) uniform generation and approximate counting arise from quite dissimilar areas. However, there is a deep connection between approximate counting and almost uniform generation. Jerrum, Valiant and Vazirani showed that for SAT the problem of generating satisfying assignments *almost uni-* *formly* is polynomially inter-reducible with *randomized approximate model counting* [14]. In [23], Stockmeyer showed that counting models within a specified tolerance factor can be achieved in deterministic polynomial time using a $\Sigma_2^p$-oracle. Building on Stockmeyer's result and the inter-reducibility of approximate counting and almost uniform generation, Jerrum et al. showed that the problem of almost uniform generation and approximate counting lie in the second level of polynomial hierarchy [14]. Our notion of approximate counting (discussed in Chapter 4) is equivalent to the notion of randomized approximate model counting used in [14].

Our work shows a similar but different close connection: in particular, we show a close connection between near-uniform generation of satisfying assignments and randomized approximate counting. In the context of generating satisfying assignments, the requirement of near uniformity is a more relaxed notion than that of almost uniformity, as proposed in [14]. However, our results does not yet generalize [14]'s result, as will be discussed in Chapter 4. In fact, our results throws open the question of whether the notions of almost uniform generation and near uniform generation are themselves inter-reducible. This is discussed further in Chapter 5. In the next section, we list and discuss the main contributions of this thesis.

## 1.4 Contributions

The main contribution of this thesis is a novel approach to solve the problems of uniform generation and approximate model counting of SAT witnesses. Our approach, which is based on limited-independence hashing, provides theoretical guarantees and also scales to formulas with thousands of variables.

We describe UniWit, a randomized algorithm, that near-uniformly samples the solution space of Boolean formulas and demonstrate its practical utility over large constraint sets. We also show that UniWit performs better than previous best-of-breed algorithms for this problem in terms of both run time and uniformity.

A novel algorithm, ApproxMC, is proposed, which to the best of our knowledge is the first scalable approximate model counter for CNF formulas. Experimental comparison over a large set of problems arising from various domains show that ApproxMC reports counts that are close to the exact counts. ApproxMC also succeeds in reporting counts with small tolerance and with high confidence in cases that are too large for computing exact model counts. Both UniWit and ApproxMC make polynomial numbers of calls to a SAT solver and run in randomized polynomial time relative to a SAT oracle.

This thesis also contributes two tools, UniWit and ApproxMC, that are based on the algorithms as described above. Both tools have been made available in the public domain.

The algorithms presented in this thesis were published in [24, 25].

## 1.5  Organization

The remainder of this thesis is organized as follows.

Chapter 2 presents notation and describes preliminaries needed for the subsequent discussion. The notion of limited-independence hashing, which is central to the work presented in this thesis, is briefly surveyed in this chapter.

Chapter 3 discusses the problem of uniform generation of SAT witnesses. It proposes a novel approach with theoretical performance guarantees and demonstrates the practical utility of the approach over an extensive set of benchmarks.

Chapter 4 presents the problem of approximate counting of SAT witnesses. It proposes the first scalable approximate model counting algorithm for CNF formu-las. The practical utility of the approach is demonstrated over an extensive set of benchmarks arising from the application domains of model counting.

Finally, Chapter 5 summarizes the main contributions of this thesis and outlines possible future directions.

# Chapter 2

# Preliminaries

In this chapter we introduce notations and preliminaries needed to present our work. We begin with some basic notations.

## 2.1 Notations

Let $\Sigma$ be an alphabet and $R \subseteq \Sigma^* \times \Sigma^*$ be a binary relation. We say that $R$ is an $\mathcal{NP}$-relation if $R$ is polynomial-time decidable, and if there exists a polynomial $p(\cdot)$ such that for every $(x,y) \in R$, we have $|y| \leq p(|x|)$. Let $L_R$ be the language $\{x \in \Sigma^* \mid \exists y \in \Sigma^*, (x,y) \in R\}$. The language $L_R$ is said to be in $\mathcal{NP}$ if $R$ is an $\mathcal{NP}$-relation. The set of all satisfiable propositional logic formulas in CNF is a language in $\mathcal{NP}$. Given $x \in L_R$, a *witness* or *model* of $x$ is a string $y \in \Sigma^*$ such that $(x,y) \in R$. The set of all models of $x$ is denoted $R_x$. For notational convenience, we fix $\Sigma$ to be $\{0,1\}$ without loss of generality. If $R$ is an $\mathcal{NP}$-relation, we may further assume that for every $x \in L_R$, every witness $y \in R_x$ is in $\{0,1\}^n$, where $n = p(|x|)$ for some polynomial $p(\cdot)$. Throughout this work, we use $\Pr[X]$ to denote the probability of outcome $X$ of sampling from a probability space while $\mathsf{E}[X]$ and $\sigma^2[X]$ denote the expectation and variance for a random variable $X$. We denote probability distribution of a set of variables $V$ by $\{_V(V)$.

### 2.1.1 Notations Related to Uniformity

Given a $\mathcal{NP}$ relation $R$, a *probabilistic generator* of witnesses for $R$ is a probabilistic algorithm $\mathscr{G}(\cdot)$ that takes as input a string $x \in L_R$ and generates a random witness of $x$. A *uniform generator* $\mathscr{G}^u(\cdot)$ is a probabilistic generator that guarantees $\Pr[\mathscr{G}^u(x) = y] = 1/|R_x|$ for every witness $y$ of $x$. An *almost uniform generator* $\mathscr{G}^{au}(\cdot)$ relaxes the guarantee of uniformity, and ensures that for every $y \in R_x$, we have $(1+\varepsilon)^{-1}\varphi(x) \leq \Pr[\mathscr{G}^{au}(x) = y] \leq (1+\varepsilon)\varphi(x)$, where $\varepsilon > 0$ is the specified tolerance and $\varphi(x)$ is an appropriate function [18, 14]. A *near-uniform generator* $\mathscr{G}^{nu}(\cdot)$ further relaxes the guarantee of uniformity, and ensures that $\Pr[\mathscr{G}^{nu}(x) = y] \geq c \cdot (1/|R_x|)$ for a constant $c$, where $0 < c \leq 1$. Clearly, the larger $c$ is, the closer a near-uniform generator is to being a uniform generator.

Like previous works [18, 14], we allow our generator to occasionally "fail", i.e. the generator may occasionaly output no witness, but a special failure symbol $\perp$. A generator that occasionally fails must have its failure probability bounded above by $d$, where $d$ is a positive constant strictly less than 1.

### 2.1.2 Notations Related to Counting

The *counting problem* corresponding to $R$ asks "Given $x \in \{0,1\}^*$, what is $|R_x|$?". If $R$ relates CNF propositional formulae to their satisfying assignments, the corresponding counting problem is called #SAT. Let $\varepsilon$ and $\delta$ be real numbers such that $0 < \varepsilon \leq 1$ and $0 < \delta \leq 1$. For every propositional formula $F$, let #$F$ denote the number of models. A *counter* of solutions of $F$ is an algorithm $\mathscr{J}(\cdot)$ that takes as input $F$ and an optional list of parameters: confidence ($\delta$) and tolerance ($\varepsilon$), and returns a count estimating #$F$. An *exact counter* $\mathscr{J}^e(\cdot)$ guarantees $\mathscr{J}^e(F) = |R_x|$. An $(\varepsilon, \delta)$ *counter* $\mathscr{J}^a(\cdot)$ guarantees that $\Pr[(1+\varepsilon)^{-1}\#F \leq \mathscr{J}^a(F, \varepsilon, \delta) \leq \#F] \geq 1 - \delta$ [26]. An upper bounding counter $\mathscr{J}^u(\cdot)$ ensures that $\Pr[\mathscr{J}^u(F,\delta) \geq \#F] \geq 1 - \delta$ [27]. Similarly a lower bounding counter $\mathscr{J}^l(\cdot)$ ensures that $\Pr[\mathscr{J}^l(F,\delta) \geq \#F] \leq 1 - \delta$. Noe that bounding counters *do not* provide any tolerance guarantees. A guarantee-less counter $\mathscr{J}^g(\cdot)$ does not provide any guarantees over the estimate of the count.

## 2.2 Standard Probability Results

We state some standard probability results that are used throughout this work. Standard textbooks [28, 29] can be consulted for detailed information.

**r-wise Independence**

A Set $V$ of random variables is said to exhibit $r-$wise independence *iff* for every subset of $V$ size $r$ or less, the joint probability distribution function of the subset is equal to product of individual marginal distributions.

**Markov Inequality**

Let $X$ be a nonnegative random variable and $a > 0$, then

$$\Pr[X > a] \leq \frac{\mathsf{E}[X]}{a} \qquad (2.1)$$

**Chebyshev Inequality**

Let $\beta > 0$, then

$$\Pr[|X - \mathsf{E}[X]| \geq \beta \sigma^2[X]] \leq \frac{1}{\beta^2} \qquad (2.2)$$

**Tail Bound for Pairwise Independent Hash Functions**

We use Chebyshev inequality to obtain a tighter tail bound for pairwise inequality.

**Theorem 2.2.1.** *Let $\Gamma$ be the sum of 2-wise independent random variables, each of which is confined to the interval $[0,1]$, and suppose $\mathsf{E}[\Gamma] = \mu$. For $0 < \beta \leq 1$, if $2 \leq \left\lfloor \beta^2 \mu e^{-1/2} \right\rfloor$, then $\Pr[|\Gamma - \mu| \geq \beta \mu] \leq e^{-3/2}$.*

*Proof.* Let $X_i, i \in [n]$ be r-wise independent random variables confined to the interval $[0,1]$ and $\Gamma = \Sigma_{i=1}^n X_i$ with $E[\Gamma] = \mu$. Let $\sigma^2[\Gamma]$ denote the variance of $\Gamma$ and for pairwise independent hash functions, we have $\sigma^2[\Gamma] = \Sigma_{i=1}^n \sigma^2[X_i]$. Since $X_i \in [0,1]$, we have $\sigma^2[X_i] \leq E[X_i]$. Thus, $\sigma^2[\Gamma] \leq E[\Gamma]$. The proof is now completed by applying Chebyshev's inequality.

$$\Pr(|\Gamma - \mathsf{E}[\Gamma]| \geq \beta \mathsf{E}[\Gamma]) \leq \frac{\sigma^2[\Gamma]}{(\beta \mathsf{E}[\Gamma])^2} \leq \frac{e^{-1/2}}{3} \leq e^{-3/2} \qquad \square$$

## 2.3  Limited-Independence Hashing

A key idea in our work for uniform generation and model counting is to use limited-independence hash functions that map strings in $\{0,1\}^n$ to $\{0,1\}^m$, for $m \leq n$. The following notion and terminology used in the context of limited-independence hashing has also been discussed in the works of [18] and [30].

**Definitions**

Let $n, m$ and $r$ be positive integers, and let $H(n,m,\cdot)$ denote a family of hash functions mapping from $\{0,1\}^n$ to $\{0,1\}^m$. We use $h \xleftarrow{R} H(n,m,\cdot)$ to denote the act of choosing a hash function $h$ uniformly at random from $H(n,m,\cdot)$. We say that a family $H(n,m,\cdot)$ exhibits $r-$wise independence if for each $\alpha_1, \ldots \alpha_r \in \{0,1\}^m$ and for each distinct $y_1, \ldots y_r \in \{0,1\}^n$, $\Pr\left[\bigwedge_{i=1}^r h(y_i) = \alpha_i : h \xleftarrow{R} H(n,m,\cdot)\right] = 2^{-mr}$. We use $H(n,m,r)$ to denote such a family of $r-$wise independent hash functions.

For every $\alpha \in \{0,1\}^m$ and $h \in H(n,m,r)$, let $h^{-1}(\alpha)$ denote the set $\{y \in \{0,1\}^n \mid h(y) = \alpha\}$. Given $R_x \subseteq \{0,1\}^n$ and $h \in H(n,m,r)$, we use $R_{x,h,\alpha}$ to denote the set $R_x \cap h^{-1}(\alpha)$. If we keep $h$ fixed and let $\alpha$ range over $\{0,1\}^m$, the sets $R_{x,h,\alpha}$ form a partition of $R_x$. Following the notation of [18], we call each element of such a partition a *cell* of $R_x$ induced by $h$. It has been shown in [18] that if $h$ is chosen uniformly at random from $H(n,m,r)$ for $r \geq 1$, the expected size of $R_{x,h,\alpha}$, denoted $\mathsf{E}\left[|R_{x,h,\alpha}|\right]$, is $|R_x|/2^m$, for each $\alpha \in \{0,1\}^m$.

**Construction of Limited-Independence Hash Functions**

In [18], the authors suggest using polynomials over finite fields to generate *r*-wise independent hash functions. We call these *algebraic* hash functions and denote by $H_{alg}(n,m,r)$. Choosing a random algebraic hash function $h \in H_{alg}(n,m,r)$ requires choosing a sequence $(a_0, \ldots a_{r-1})$ of elements in the field $\mathbb{F} = \mathrm{GF}(2^{\max(n,m)})$, where $GF(2^k)$ denotes the Galois field of $2^k$ elements. Given $y \in \{0,1\}^n$, the hash value $h(y)$ can be computed by interpreting $y$ as an element of $\mathbb{F}$, computing $\Sigma_{j=0}^{r-1} a_j y^j$ in $\mathbb{F}$, and selecting $m$ bits of the encoding of the result. Unfortunately it is well known that the multiplier operator for Galois field is quite expensive [31], thus making this approach impractical.

**Efficient Limited-Independence Hash Functions**

Our approach uses computationally efficient linear hash functions. In particular, we use pairwise and 3-wise independent hash functions. The literature describes several families of efficiently computable pairwise independent hash functions. One such family, which we denote $H_{conv}(n,m,2)$, is based on the *wrapped convolution* function [32]. For $a \in \{0,1\}^{n+m-1}$ and $y \in \{0,1\}^n$, the wrapped convolution $c = (a \bullet y)$ is defined as an element of $\{0,1\}^m$ as follows: for each $i \in \{1,\ldots m\}$, $c[i] = \bigoplus_{j=1}^{n}(y[j] \wedge a[i+j-1])$, where $\bigoplus$ denotes logical xor and $v[i]$ denotes the $i^{th}$ component of the bit-vector $v$. The family $H_{conv}(n,m,2)$ is defined as $\{h_{a,b}(y) = (a \bullet y) \oplus_m b \mid a \in \{0,1\}^{n+m-1}, b \in \{0,1\}^m\}$, where $\oplus_m$ denotes componentwise xor of two elements of $\{0,1\}^m$. By randomly choosing $a$ and $b$, we can randomly choose a function $h_{a,b}(x)$ from this family. It has been shown in [32] that $H_{conv}(n,m,2)$ is pairwise independent.

Another computationally efficient family, which we denote $H_{xor}(n,m,3)$, is based on randomly choosing bits from $y \in \{0,1\}^n$ and xor-ing them. This family of hash functions has been used in earlier works [30], and has been shown to be 3-independent (therefore pairwise independent as well) in [33]. Let $h(y)[i]$ denote the $i^{th}$ component of the bit-vector obtained by applying hash function $h$ to $y$. The family $H_{xor}(n,m,3)$ is defined as $\{h(y) \mid (h(y))[i] = a_{i,0} \oplus (\bigoplus_{k=1}^{n} a_{i,k} \cdot y[k]), a_{i,j} \in \{0,1\}, 1 \leq i \leq m, 0 \leq j \leq n\}$. By randomly choosing the $a_{i,j}$'s, we can randomly choose a hash function from this family.

The algorithms presented in this thesis work with any pairwise independent family of hash functions. The reference implementation and experimental analysis of our algorithms uses the above two families of hash functions.

# Chapter 3

# Uniform Generation

We use "uniform generation" to refer to the problem of generating satisfying assignments of a propositional formula near-uniformly at random from the space of all satisfying assignments. As mentioned in Chapter 1, this is an important problem with applications to wide variety of applications areas ranging from random directed testing to probabilistic reasoning [12, 13]. Since a significant body of constraints that arise in testing and verification settings and in other application areas (like probabilistic reasoning) can be efficiently encoded as Boolean constraints in conjunctive normal form (CNF), we focus on the problem of uniform generation of satisfying assignments of CNF formulas. Following terminology used in the literature, such assignments are henceforth called *SAT Witnesses*.

The problem of uniform generation of CNF formulas has been of long-standing theoretical and practical interest [34, 23, 33]. Industrial approaches to solving this problem either rely on Reduced Ordered Binary Decision Diagrams(ROBDD)-based techniques [11] , which do not scale well (see, for example, the comparison in [35]), or use heuristics that offer no guarantee on performance or uniformity when applied to large problem instances[*]. Prior published work in this area broadly belong to one of two categories. In the first category, the focus is on heuristic sampling techniques that scale to large systems of constraints [36, 37, 33, 35]. Markov Chain Monte Carlo (MCMC) methods and techniques based on random seedings of SAT solvers belong to this category. These methods, however, either offer very weak or no guarantees on the uniformity of sampling (see [35] for a comparison), or require the user to provide hard-to-estimate problem-specific parameters that crucially affect the performance and uniformity of sampling. In the second category of work, the focus is on stronger guarantees on uniformity of sampling [18, 14, 11]. Unfortunately, our experience indicates that these techniques do not scale even to relatively small problem instances (involving few tens of variables) in practice.

---

[*]Private communication: R. Kurshan

The work presented here tries to bridge the above mentioned extremes. Specifically, we provide guarantees of near-uniform sampling, and of a bounded probability of failure, without the user having to provide any hard-to-estimate parameters. We also demonstrate that our proposed approach scales in practice to constraints involving thousands of variables. Note that there is evidence that uniform generation of SAT witnesses is harder than SAT solving [14]. Thus, while today's SAT solvers are able to handle hundreds of thousands of variables and more, we believe that scalability of our approach to thousands of variables should be considered a major improvement in this area.

The remainder of this Chapter is organized as follows. In Section 3.1, we review previous works in this area. Design choices behind our algorithm and some implementation issues are discussed in Section 3.2. A mathematical analysis of the guarantees provided by our algorithm is presented in Section 3.3. Section 3.4 discusses experimental results on a large set of benchmarks. Our experiments demonstrate that our algorithm is more efficient in practice and generates witnesses that are more evenly distributed than those generated by the best known alternative algorithm that scales to comparable problem sizes.

## 3.1 Prior Work

In this section, we briefly review various approaches proposed in the literature for uniform generation in SAT and related domains. Finally we discuss in detail two algorithms that come closest to our work.

**Markov Chain Monte Carlo (MCMC)-based Methods:** A wide variety of MCMC-based algorithms have been proposed in the literature to sample from complex distributions. These include Metropolis algorithm, simulated annealing and the [37, 38, 39]. The core idea of these algorithms is to sample using carefully chosen Markov chains in which the steady state distribution matches the desired distribution. MCMC methods guarantee convergence to uniform distribution only when run for sufficiently long time. Most practical algorithms based on MCMC methods, however, use heuristic adaptations to ensure better performance. For example, Wei, Erenrich and Selman proposed an algorithm, named SampleSAT, based on a hybrid strategy involving random walks and simulated annealing [40]. Kitchen and Kuehlmann [35] proposed an MCMC based approach using Metropolis-Hasting sampler to generate stimuli for Boolean/integer constraint problems. Unfortunately, the adaptations used in the above algorithm fail to provide any guarantees of uniformity.

**Weighted Binary Decision Diagram (BDD) based Methods:** A new approach based on sampling from a set of constraints based on weighted binary decision diagrams [41] was proposed in [42, 43]. The core idea of the algorithm is to construct a BDD-based on the input constraints and then generate uniform samples in a single pass over the BDD. The approach works well for small to medium-sized examples but does not scale to larger problems. Hence it is not scalable to large problems in practice. A detailed analysis of the scalability limitations of BDD-based methods is presented in [35].

An alternative approach to uniform generation based on BDDs was proposed in [44]. This approach relies on constructing an equivalent circuit for BDD constraints [44]. Unfortunately, this approach fails to provide guarantees of uniformity [35].

**Interval-propagation-based sampling:** Interval-propagation-based sampling techniques have been used by some researchers to address the scalability challenges posed by uniform generation in practice [45]. The central idea underlying these techniques is to maintain intervals of values that a variable can take and generate samples by performing random sampling over these intervals. The simplicity of such approaches provides good performance in practice but the distributions generated can deviate significantly from the uniform distribution [35].

**Belief networks:** Another class of methods based on Constrained satisfaction problems (CSP), particularly belief propagation, have been proposed in [46, 47]. The proposed techniques improve on the traditional MCMC based methods by integrating sampling with back-jumping search and no-good learning. Experimental comparisons, however, have shown that these techniques perform poorly compared to MCMC based techniques with random walk and simulated annealing heuristics, as in SampleSAT [47].

**Hashing-based techniques** Sipser poinneered hashing-based approach in [34] building upon the universal hashing introduced by Carter and Wegman [48]. This has subsequently been used in theoretical [18] and practical [33] treatments of uniform sampling. The key idea in these works is to randomly partition the solution space into "small cells" of *roughly* equal size. The act of picking a solution randomly chosen cell provides the required guarantees. Our work also falls in this category, however there are notable differences as discussed below.

### 3.1.1 BGP and XORSample′ Algorithms

We now discuss two algorithms that come closest to our work. In 1998, Bellare, Goldreich, and Petrank proposed an algorithm, showing that uniform generation of $\mathcal{NP}$-witnesses can be achieved in probabilistic polynomial time using an $\mathcal{NP}$-oracle [18]. This improved on previous work by Jerrum, Valiant and Vazirani [14], who showed that uniform generation can be achieved in probabilistic polynomial time using a $\Sigma_2^P$ oracle, and almost-uniform generation can be achieved in probabilistic polynomial time using an $\mathcal{NP}$ oracle. In the remainder of this chapter, we refer to Bellare et al.'s algorithm as the BGP algorithm (after the last names of the authors).

Let $R$ be an $\mathcal{NP}$-relation over $\Sigma$. The BGP algorithm takes as input an $x \in L_R$ and either generates a witness that is uniformly distributed in $R_x$, or produces a symbol $\perp$ (indicating a failed run). The pseudo-code for BGP is presented in Algorithm 1. In the presentation, we assume w.l.o.g. that $n$ is an integer such that $R_x \subseteq \{0,1\}^n$. We also assume access to $\mathcal{NP}$-oracles to answer queries about cardinalities of witness sets and also to enumerate small witness sets.

For clarity of exposition, we have made a small adaptation to the algorithm originally presented in [18]. Specifically, if $h$ does not satisfy $(\forall \alpha \in \{0,1\}^{i-l}, |R_{x,h,\alpha}| \leq 2n^2)$ when the loop in lines 7–10 terminates, the original algorithm forces a specific choice of $h$. Instead, algorithm BGP simply outputs $\perp$ (indicating a failed run) in this situation. A closer look at the analysis presented in [18] shows that all results continue to hold with this adaptation. The authors of [18] use algebraic hash functions and random choices of $n$-tuples in $\text{GF}(2^{\max(n,i-l)})$ to implement the selection of a random hash function in line 9 of the pseudocode. The following theorem summarizes the key properties of the BGP algorithm [18].

---

**Algorithm 1** BGP($x$) :

/* Assume $R_x \subseteq \{0,1\}^n$ */

1: pivot $\leftarrow 2n^2$;

2: **if** $|R_x| \leq$ pivot **then**

3:     List all elements $y_1, \ldots y_{|R_x|}$ of $R_x$;

4:     Choose $j$ at random from $\{1, \ldots |R_x|\}$, and **return** $y_j$;

5: **else**

6:     $l \leftarrow 2\lceil \log_2 n \rceil$; $i \leftarrow l$;

7:     **repeat**

8:         $i \leftarrow i+1$;

9:         Choose $h$ at random from $H_{alg}(n, i-l, n)$;

10:    **until** ($\forall \alpha \in \{0,1\}^{i-l}, |R_{x,h,\alpha}| \leq 2n^2$) or ($i = n-1$);

11:    **if** ($\exists \alpha \in \{0,1\}^{i-l}, |R_{x,h,\alpha}| > 2n^2$) **then return** $\bot$;

12:    **else**

13:        Choose $\alpha$ at random from $\{0,1\}^{i-l}$;

14:        List all elements $y_1, \ldots y_{|R_{x,h,\alpha}|}$ of $R_{x,h,\alpha}$;

15:        Choose $j$ at random from $\{1, \ldots \text{pivot}\}$;

16:        **if** $j \leq |R_{x,h,\alpha}|$ **then return** $y_j$;

17:        **else return** $\bot$;

---

**Theorem 3.1.1.** *The output generated by* BGP *is uniformly distributed. Formally, if a run of the* BGP *algorithm is successful, the probability that* $y \in R_x$ *is* $1/|R_x|$. *Further, the probability that a run of the algorithm fails is* $\leq 0.8$.

Since the probability of any witness $y \in R_x$ being output by a successful run of the algorithm is independent of $y$, the BGP algorithm guarantees uniform generation of witnesses. However, as we argue in the next section, scaling the algorithm to even medium-sized problem instances is quite difficult in practice. Indeed, we have found no published report discussing any implementation of the BGP algorithm.

In 2007, Gomes, Sabharwal and Selman [33] presented two closely related algorithms named XORSample and XORSample' for near-uniform sampling of combinatorial spaces. A key idea in both these algorithms is to constrain a given instance $F$ of the CNF SAT problem by a set of randomly selected *xor constraints* over the variables appearing in $F$. A xor constraint over a set $V$ of variables is an equation of the form $e = c$, where $c \in \{0, 1\}$ and $e$ is the logical xor of a subset of $V$. A probability distribution $\mathbb{X}(|V|, q)$ over the set of all xor constraints over $V$ is characterized by the probability $q$ of choosing a variable in $V$. A random xor constraint from $\mathbb{X}(|V|, q)$ is obtained by forming a xor constraint where each variable in $V$ is chosen independently with probability $q$, and $c$ is chosen uniformly at random.

We present the pseudo-code of algorithm XORSample' below. The algorithm uses a function SATModelCount that takes a Boolean formula $F$ and returns the exact count of witnesses of $F$. Algorithm XORSample' takes as inputs a CNF formula $F$, the parameter $q$ discussed above and an integer $s > 0$. Suppose the number of variables in $F$ is $n$. The algorithm proceeds by conjoining $s$ xor constraints to $F$, where the constraints are chosen randomly from the distribution $\mathbb{X}(n, q)$. Let $F'$ denote the conjunction of $F$ and the random xor constraints, and let $mc$ denote the model count (i.e., number of witnesses) of $F'$. If $mc \geq 1$, the algorithm enumerates the witnesses of $F'$ and chooses one witness at random. Otherwise, the algorithm outputs $\bot$, indicating a failed run. Algorithm XORSample can be viewed as a variant of algorithm

---

**Algorithm 2** XORSample$'(F,q,s)$
---
/* $n =$ Number of variables in $F$ */

1: $Q_s \leftarrow \{s$ random xor constraints from $\mathbb{X}(n,q)\}$;

2: $F' = F \wedge (\bigwedge_{f \in Q_s} f)$;

3: $mc \leftarrow$ SATModelCount$(F')$;

4: **if then**$(mc \geq 1)$

5:     Choose $i$ at random from $\{1, \ldots mc\}$;

6:     Enumerate the first $i$ witnesses of $F'$;

7:     **return** $i^{th}$ witness of $F'$;

8: **else return** $\bot$;

---

XORSample$'$ in which we check if $mc$ is exactly 1 (instead of $mc \geq 1$) in line 4 of the pseudocode. An additional difference is that if the check in line 4 fails, algorithm XORSample starts afresh from line 1 by randomly choosing $s$ xor constraints. In our experiments, we observed that XORSample$'$ significantly outperforms XORSample in performance, hence we consider only XORSample$'$ for comparison with our algorithm.

Let $\langle f_1, \ldots f_s \rangle$ denote the lexicographic ordering of the random xor constraints in $Q_s$. Choosing the set $Q_s$ is equivalent to choosing a random hash function $h_{Q_s} : \{0,1\}^n \rightarrow \{0,1\}^s$, where $h_{Q_s}[i] = f_i$ for $i \in \{1, \ldots s\}$. In [33], the authors showed that if $q = \frac{1}{2}$, the random hash function $h_{Q_s}$ is 3-wise independent, i.e. in $H_{xor}(n,s,3)$. This property was subsequently used in [33] to prove the following key theorem.

**Theorem 3.1.2.** *Let $F$ be a SAT formula with $2^{s^*}$ solutions. Let $\alpha > 0$ and $s = s^* - \alpha$. For a witness $y$ of $F$, the probability with which* XORSample′ *with parameters $q = \frac{1}{2}$ and $s$ outputs $y$ is bounded below by $c'(\alpha)2^{-s^*}$, where $c'(\alpha) = \frac{1-2^{-\alpha/3}}{(1+2^{-\alpha})(1+2^{-\alpha/3})}$. Further,* XORSample′ *succeeds with probability larger than $c'(\alpha)$.*

Since the performance of the algorithm crucially depeneds on the input parameter $\alpha$, a good estimate of $s^*$ is needed. Finding number of solutions of a SAT formula $F$, however, is #$P$−complete and therefore,authors of [33] propose a binary search heuristic to estimate $s^*$. The search heuristic, however, is computationally very expensive as demonstrated by our experimental results in Section 3.5. While the choice of $q = \frac{1}{2}$ allowed the authors of [33] to prove Theorem 3.1.2, the authors acknowledge that finding witnesses of $F'$ is quite hard in practice when random xor constraints are chosen from $\mathbb{X}(n, \frac{1}{2})$. Therefore, they advocate using values of $q$ much smaller than $\frac{1}{2}$. Unfortunately, the analysis that yields the theoretical guarantees in Theorem 3.1.2 does not hold with these smaller values of $q$ [49]. This illustrates the conflict between witness generators with good performance in practice, and those with good theoretical guarantees.

## 3.2 The UniWit Algorithm

We now describe an adaptation, called UniWit, of the BGP algorithm that scales to much larger problem sizes than those that can be handled by the BGP algorithm, while weakening the guarantee of uniform generation to that of near-uniform generation. Experimental results, however, indicate that the witnesses generated by our algorithm are quite uniform in practice. Our algorithm can also be viewed as an adaptation of the XORSample′ algorithm, in which we do not need to provide hard-to-estimate problem-specific parameters like $s$ and $q$.

We begin with some observations about the BGP algorithm. In what follows, line numbers refer to those in the pseudo-code of the BGP algorithm presented in Section 3.1.1. Our first observation is that the loop in lines 7–10 of the pseudo-code iterates until either $|R_{x,h,\alpha}| \leq 2n^2$ for *every* $\alpha \in \{0,1\}^{i-l}$ or $i$ increments to $n-1$. Checking the first condition is computationally prohibitive even for values of $i-l$ and $n$ as small as few tens. So we ask if this condition can be simplified, perhaps with some weakening of theoretical guarantees. Indeed, we have found that if the condition requires that $1 \leq |R_{x,h,\alpha}| \leq 2n^2$ for a *random* $\alpha \in \{0,1\}^{i-l}$ (instead of for *every* $\alpha \in \{0,1\}^{i-l}$), we can still guarantee near-uniformity (but not uniformity) of the generated witnesses. This suggests choosing both a random $h \in H(n, i-l, n)$ and a random $\alpha \in \{0,1\}^{i-l}$ within the loop of lines 7–10.

The analysis presented in [18] relies on $h$ being sampled uniformly from a family of $n$-wise independent hash functions. In the context of generating SAT witnesses, $n$ denotes the number of propositional variables in the input formula. This can be large (several thousands) in problems arising from directed random testing. Unfortunately, implementing $n$-wise independent hash functions using algebraic hash functions (as advocated in [18]) for large values of $n$ is computationally infeasible in practice. This prompts us to ask if the BGP algorithm can be adapted to work with $r$-wise independent hash functions for small values of $r$, and if simpler families of hash functions can be used. Indeed, we have found that with $r \geq 2$, an adapted version of the BGP algorithm can be made to generate near-uniform witnesses. We can also bound the probability of failure of the adapted algorithm by a constant. The sufficiency of pairwise independence allows us to use computationally efficient xor-based families of hash functions, like $H_{conv}(n,m,2)$ discussed in discussed in Section 2.3. This provides a significant scaling advantage to our algorithm vis-a-vis the BGP algorithm in practice.

In the context of uniform generation of SAT witnesses, checking if $|R_x| \leq 2n^2$ (line 2 of pseudo-code) or if $|R_{x,h,\alpha}| \leq 2n^2$ (line 10 of pseudo-code, modified as suggested above) can be done either by approximate model-counting or by repeated invocations of a SAT solver. State-of-the-art approximate model counting techniques rely on randomly sampling the witness space, suggesting a circular dependency [30]. Hence, we choose to use a SAT solver as the back-end engine for enumerating and counting witnesses. Note that if $h$ is chosen randomly from $H_{conv}(n,m,2)$, the formula for which we seek witnesses is the conjunction of the original (CNF) formula and xor constraints encoding the inclusion of each witness in $h^{-1}(\alpha)$. We therefore choose to use a SAT solver optimized for conjunctions of xor constraints and CNF clauses as the back-end engine; specifically, we use CryptoMiniSAT (version 2.9.2) [50].

Modern SAT solvers often produce partial assignments that specify values of a subset of variables, such that every assignment of values to the remaining variables gives a witness. Since we must find large numbers of witnesses($2n^2 \approx 2 \times 10^6$ if $n \approx 1000$), it would be useful to obtain partial assignments from the SAT solver. Unfortunately, conjoining random xor constraints to the original formula reduces the likelihood that large sets of witnesses can be encoded as partial assignments. Thus, each invocation of the SAT solver is likely to generate only a few witnesses, necessitating a large number of calls to the solver. To make matters worse, if the count of witnesses exceeds $2n^2$ and if $i < n-1$, the check in line 10 of the pseudo-code of algorithm BGP (modified as suggested above) fails, and the loop of lines 7–10 iterates once more, requiring generation of up to $2n^2$ witnesses of a modified SAT problem all over again. This can be computationally prohibitive in practice. Indeed, our implementation of the BGP algorithm with CryptoMiniSAT failed to terminate on formulas with few tens of variables, even when running on high-performance computers for 20 hours. This prompts us to ask if the required number of witnesses, or *pivot*, in the BGP algorithm (see line 1 of the pseudo-code) can be reduced. We answer this question in the affirmative, and show that the pivot can indeed be reduced to $2n^{1/k}$, where $k$ is an integer $\geq 1$. Note that if $k = 3$ and $n = 1000$, the value of $2n^{1/k}$ is only 20, while $2n^2$ equals $2 \times 10^6$. This translates to a significant improvement in the sizes of problems for which we can generate random witnesses. There are, however, some practical tradeoffs involved in the choice of $k$; we defer a discussion of these to a later part of this section.

In lines 13–16, the value of $j$ is chosen from the set of $\{1, \ldots \ldots \text{pivot}\}$ instead of $\{1, \ldots \ldots |R_{x,h,\alpha}|\}$. This allows authors to obtain stronger guarantees of uniformity while weakening success probability slightly. We continue to use this insight in our algoirthm.

We now present the UniWit algorithm, which implements the proposed modifications to the BGP algorithm. UniWit takes as inputs a CNF formula $F$ with $n$ variables, and an integer $k \geq 1$. The algorithm either outputs a witness that is near-uniformly distributed over the space of all witnesses of $F$, or produces a symbol $\perp$ indicating a failed run. We also assume that we have access to a function BoundedSAT that takes as inputs a propositional formula $F$ that is a conjunction of a CNF formula and xor constraints, and an integer $r \geq 0$, and returns a set $S$ of witnesses of $F$ such that $|S| = \min(r, \#F)$, where $\#F$ denotes the count of all witnesses of $F$.

**Implementation issues:** There are four steps in UniWit (lines 4, 9, 10 and 16 of the pseudo-code) where random choices are made. In our implementation, in line 10 of the pseudo-code, we choose a random hash function from the family $H_{conv}(n, i-l, 2)$, since it is computationally efficient to do so. Recall from Section 2.3 that choosing a random hash function from $H_{conv}(n,m,2)$ requires choosing two random bit-vectors. It is straightforward to implement these choices and also the choice of a random $\alpha \in \{0,1\}^{i-l}$ in line 10 of the pseudo-code, if we have access to a source of independent and uniformly distributed random bits. In lines 4 and 16, we must choose a random integer from a specified range. By using standard techniques (see, for example, the discussion on coin tossing in [18]), this can also be implemented efficiently if we have access to a source of random bits. Our implementation uses random sequences of bits generated from nuclear decay processes and available at HotBits [51]. We download and store a long sequence of random bits in a file ( 12 MB), and access an appropriate number of bits sequentially, with wrap around, whenever needed. We defer experimenting with sequences of bits obtained from other pseudo-random generators to a future study.

**Algorithm 3** UniWit($F,k$)
　　/*Assume $z_1,\cdots,z_n$ are variables in $F$ */
1: pivot $\leftarrow \lceil 2n^{1/k} \rceil$; $S \leftarrow$ BoundedSAT($F$, pivot $+1$);
2: **if** $|S| \leq$ pivot **then**
3: 　　Let $y_1,\ldots y_{|S|}$ be the elements of $S$;
4: 　　Choose $j$ at random from $\{1,\ldots |S|\}$ and **return** $y_j$;
5: **else**
6: 　　$l \leftarrow \lfloor \frac{1}{k} \cdot (\log_2 n) \rfloor$; $i \leftarrow l$;
7: 　　**repeat**
8: 　　　　$i \leftarrow i+1$;
9: 　　　　Choose $h$ at random from $H_{conv}(n,i-l,2)$;
10: 　　　　Choose $\alpha$ at random from $\{0,1\}^{i-l}$;
11: 　　　　$S \leftarrow$ BoundedSAT($F \wedge (h(z_1,\ldots z_n) = \alpha)$, pivot $+1$);
12: 　　**until** $(1 \leq |S| \leq$ pivot$)$ or $(i = n)$;
13: 　　**if** $(|S| >$ pivot$)$ or $(|S| < 1)$ **then return** $\bot$;
14: 　　**else**
15: 　　　　Let $y_1,\ldots y_{|S|}$ be the elements of $S$;
16: 　　　　Choose $j$ at random from $\{1,\ldots pivot\}$;
17: 　　　　**if** $j \leq |S|$ **then return** $y_j$
18: 　　　　**else return** $\bot$;

In line 11 of the pseudo-code for UniWit, we invoke BoundedSAT with arguments $F \wedge (h(z_1, \ldots z_n) = \alpha)$ and pivot $+ 1$. The function BoundedSAT is implemented using CryptoMiniSAT (version 2.9.2), which allows passing a parameter indicating the maximum number of witnesses to be generated. The sub-formula $(h(z_1, \ldots z_n) = \alpha)$ is constructed as follows. As mentioned in Section 2.3, a random hash function from the family $H_{conv}(n, i-l, 2)$ can be implemented by choosing a random $a \in \{0,1\}^{n+i-l-1}$ and a random $b \in \{0,1\}^{i-l}$. Recalling the definition of $h$ from Section 2.3, the sub-formula $(h(z_1, \ldots z_n) = \alpha)$ is given by $\bigwedge_{j=1}^{i-l} \quad \bigoplus_{p=1}^{n}(z_p \wedge a[j+p-1]) \oplus b[j] \Leftrightarrow \alpha[j]$ .

**A heuristic optimization:** A (near-)uniform generator is likely to be invoked a large number of times for the same formula $F$ when generating a set of witnesses of $F$. If the performance of the generator is sensitive to problem-specific parameter(s) not known a priori, a natural optimization is to estimate values of these parameter(s), perhaps using computationally expensive techniques, in the first few runs of the generator, and then re-use these estimates in subsequent runs on the same problem instance. Of course, this optimization works only if the parameter(s) under consideration can be reasonably estimated from the first few runs. We call this heuristic optimization "*leapfrogging*".

In the case of algorithm UniWit, the loop in lines 7–12 of the pseudo-code starts with $i$ set to $l-1$ and iterates until either $i$ increments to $n$, or $|R_{F,h,\alpha}|$ becomes no larger than $2n^{1/k}$. For each problem instance $F$, we propose to estimate a lower bound of the value of $i$ when the loop terminates, from the first few runs of UniWit on $F$. In all subsequent runs of UniWit on $F$, we propose to start iterating through the loop with $i$ set to this lower bound. We call this specific heuristic *"leapfrogging i"* in the context of UniWit. Our analysis of UniWit shows that the probabilistic guarantees of UniWit continue to hold as long as the lower bound of $i$ used in leapfrogging is smaller than $\log_2 |R_F| - (1/k) \log_2 n$. The heuristic "leapfrogging $i$" however does not provide any guarantees on the lower bound of $i$, therefore the current guarantees can not be shown to hold for all inputs. Note that leapfrogging may also be used for the parameter $s$ in algorithms XORSample′ and XORSample (see pseudo-code of XORSample′). We discuss more about this in Section 3.4.

## 3.3 Analysis of UniWit

Let $R_F$ denote the set of witnesses of the input formula $F$. Using notation discussed in Section 2.1, suppose $R_F \subseteq \{0,1\}^n$. For simplicity of exposition, we assume that $\log_2 |R_F| - (1/k) \cdot \log_2 n$ is an integer in the following discussion. A more careful analysis removes this assumption with constant factor reductions in the probability of generation of an arbitrary witness and in the probability of failure of UniWit.

### 3.3.1 Near Uniformity

**Theorem 3.3.1.** *Suppose $F$ has $n$ variables, $k \geq 1$, and $n > 2^k$. For every witness $y$ of $F$, the probability that algorithm UniWit outputs $y$ on inputs $F$ and $k$ is bounded below by $\frac{1}{8|R_F|}$.*

*Proof.* Referring to the pseudo-code of UniWit, if $|R_F| \leq 2n^{1/k}$, the theorem holds trivially. Suppose $|R_F| > 2n^{1/k}$, and let $Y$ denote the event that witness $y$ in $R_F$ is output by UniWit on inputs $F$ and $k$. We break the event $Y$ into two stages: (i) termination of loop in lines 7–12 with $y$ in $R_{F,h,\alpha}$, and (ii) choosing $y$ when $y \in S$ in lines 15–17. Let $p_{i,y}$ denote the probability that the loop in lines 7–12 of the pseudo-code terminates in iteration $i$ with $y$ in $R_{F,h,\alpha}$, where $\alpha \in \{0,1\}^{i-l}$ is the value chosen in line 10. Let $p^s(y)$ denote the probability of returning $y$ when $y \in S$ in lines 15–17. It follows from the pseudo-code that $\Pr[Y] \geq p_{i,y} p^s(y)$, for every $i \in \{l, \ldots n\}$. Since $\mathsf{pivot} = 2n^{1/k}$, we have $p^s(y) = \frac{1}{\mathsf{pivot}} = \frac{1}{2n^{1/k}}$. Let us denote $\log_2 |R_F| - (1/k) \cdot \log_2 n$ by $m$, which by our assumption is an integer. Therefore, $2^m \cdot n^{1/k} = |R_F|$. Since $2n^{1/k} < |R_F| \leq 2^n$ and since $l = \lfloor (1/k) \cdot \log_2 n \rfloor$ (see line 6 of pseudo-code), we have $l < m+l \leq n$. From Lemma 3.3.2, we know that $p_{m+l,y} \geq \frac{1-n^{-1/k}}{2^{m+1}}$. Consequently, $\Pr[Y] \geq p_{m+l,y} \cdot p^s(y) \geq \frac{1-n^{-1/k}}{2^{m+2} \cdot n^{1/k}} = \frac{1-n^{-1/k}}{4|R_F|} \geq \frac{1}{8|R_F|}$, if $n > 2^k$. □

**Lemma 3.3.2.** $p_{m+l,y} \geq \frac{1-n^{-1/k}}{2^{m+1}}$

*Proof.* To calculate $p_{m+l,y}$, we first note that since $y \in R_F$, the requirement "$y \in R_{F,h,\alpha}$" reduces to "$y \in h^{-1}(\alpha)$". For $\alpha \in \{0,1\}^m$ and $y \in \{0,1\}^n$, we define $q_{m+l,y,\alpha}$ as $\Pr\left[ |R_{F,h,\alpha}| \leq 2n^{1/k} \text{ and } h(y) = \alpha : h \xleftarrow{R} H(n,m,r) \right]$, where $r \geq 2$. The proof is now completed by showing that $q_{m+l,y,\alpha} \geq (1 - n^{-1/k})/2^{m+1}$ for every $\alpha \in \{0,1\}^m$ and $y \in \{0,1\}^n$. Towards this end, we define an indicator variable $\gamma_{y,\alpha}$ for every $y \in \{0,1\}^n$ and $\alpha \in \{0,1\}^m$ as follows: $\gamma_{y,\alpha} = 1$ if $h(y) = \alpha$ and $\gamma_{y,\alpha} = 0$ otherwise. Thus, $\gamma_{y,\alpha}$ is a random variable with probability distribution induced by that of $h$. It is easy to show that (i) $\mathsf{E}[\gamma_{y,\alpha}] = 2^{-m}$, and (ii) the pairwise independence of $h$ implies pairwise independence of the $\gamma_{y,\alpha}$ variables. We now define $\Gamma_\alpha = \sum_{z \in R_F} \gamma_{z,\alpha}$ and $\mu_{y,\alpha} = \mathsf{E}[\Gamma_\alpha \mid \gamma_{y,\alpha} = 1]$. Clearly, $\Gamma_\alpha = |R_{F,h,\alpha}|$ and $\mu_{y,\alpha} = \mathsf{E}\left[\sum_{z \in R_F} \gamma_{z,\alpha} \mid \gamma_{y,\alpha} = 1\right] = \sum_{z \in R_F} \mathsf{E}[\gamma_{z,\alpha} \mid \gamma_{y,\alpha} = 1]$. Using pairwise independence of the $\gamma_{y,\alpha}$ variables, the above simplifies to $\mu_{y,\alpha} = 2^{-m}(|R_F| - 1) + 1 \leq 2^{-m}|R_F| + 1 = n^{1/k} + 1$. From Markov's inequality, we know that $\Pr[\Gamma_\alpha \leq \kappa \cdot \mu_{y,\alpha} \mid \gamma_{y,\alpha} = 1] \geq 1 - 1/\kappa$ for $\kappa > 0$. With $\kappa = \frac{2}{1 + n^{-1/k}}$, this gives $\Pr\left[|R_{F,h,\alpha}| \leq 2n^{1/k} \mid \gamma_{y,\alpha} = 1\right] \geq (1 - n^{-1/k})/2$. Since $h$ is chosen at random from $H(n, m, r)$, we also have $\Pr[h(y) = \alpha] = 1/2^m$. It follows that $q_{m+l,y,\alpha} \geq (1 - n^{-1/k})/2^{m+1}$. □

### 3.3.2 Success Probability

**Theorem 3.3.3.** *Assuming $n > 2^k$, algorithm UniWit succeeds (i.e. does not return $\bot$) with probability at least $\frac{1}{8}$.*

*Proof.* Let $P_{succ}$ denote the probability that a run of algorithm UniWit succeeds. By definition, $P_{succ} = \sum_{y \in R_F} \Pr[Y]$. Using Theorem 3.3.1, $P_{succ} \geq \sum_{y \in R_F} \frac{1}{8|R_F|} = \frac{1}{8}$. □

### 3.3.3 Complexity

**Theorem 3.3.4.** *Given an oracle for SAT, UniWit (F,k) runs in time polynomial in $|F|$ and $n^{1+1/k}$ relative to a SAT oracle where $n$ denotes the number of propositional variables in $|F|$.*

*Proof.* The pseudo-code for UniWit can be partitioned into three regions for ease of analysis: (i) line 1, (ii) line 2–4 and line 13-18, and (iii) line 7–12 (repeat-until loop). Line 1 makes one call to BoundedSAT. Lines 2–4 and 13-18 enumerate up to *pivot* solutions (each of which has length $n$), therefore take time no more than a polynomial in $n$ and *pivot*, which is in $\mathcal{O}(n^{1+1/k})$.

Referring to the pseudo-code for UniWit, the repeat-until loop is repeated $\mathcal{O}(n)$ times. Each iteration of the loop makes one call to BoundedSAT. Each call to BoundedSAT can be implemented by at most $pivot + 1$ calls to a SAT oracle, and since *pivot* is in $\mathcal{O}(n^{1/k})$, the cumulative number of calls to the SAT oracle ins all calls to BoundedSAT is in $\mathcal{O}(n^{1+1/k})$. Construction and writing of $F \wedge (h(z_1, \ldots z_n) = \alpha)$ on the memory takes time polynomial in $|F|$. Therefore, the total time taken by all calls to BoundedSAT is bounded by a polynomial in $|F|$ and $n^{1/k}$. Hence the repeat-loop in lines 7–12 takes time polynomial in $|F|$ and $n^{1+1/k}$ relative to a SAT oracle.

Summing up for all three regions, UniWit runs in time polynomial in $|F|$ and $n^{1+1/k}$ relative to a SAT oracle. □

The complexity analysis above is very conservative. In practice, we observe that the repeat-until loop iterates approximately $\log|R_F| - \log pivot$ times. This observation forms the basis of our extension of this technique to model counting (see Chapter 4). Also, the use of "leapfrogging" reduces the number of iterations of the for repeat-until loop significantly, as demonstrated by our extensive experiments (see Section 3.5).

### 3.3.4 Choice of parameter $k$

One might be tempted to use large values of the parameter $k$ to keep the value of *pivot* low. There are, however, tradeoffs involved in the choice of $k$. As $k$ increases, the pivot $2n^{1/k}$ decreases, and the chances that BoundedSAT finds more than $2n^{1/k}$ witnesses increases, necessitating further iterations of the loop in lines 7–12 of the pseudo-code. Of course, reducing the pivot also implies that BoundedSAT has to find fewer witnesses, and each invocation of BoundedSAT is likely to take less time.

The increase in the number of invocations of BoundedSAT, however, contributes to increased overall time. In our experiments, we have found that choosing $k$ to be either 2 or 3 works well for all our benchmarks (including those containing several thousand variables).

## 3.4 Experimental Results

To evaluate the performance of UniWit, we built a prototype implementation in Python and conducted an extensive set of experiments. Since our motivation stems primarily from functional verification, our benchmarks were mostly derived from functional verification of hardware designs. Specifically, we used "bit-blasted" versions of word-level constraints arising from bounded model checking of public-domain and proprietary word-level VHDL designs. In addition, we also used bit-blasted versions of several SMTLib [52, 53] benchmarks of the "QF_BV/bruttomesso/ simple_processor/" category, and benchmarks arising from "Type I" representations of ISCAS'85 circuits, as described in [54].

Our experiments were conducted on a high-performance computing cluster. Each individual experiment was run on a single node of the cluster, and the cluster allowed multiple experiments to run in parallel. Every node in the cluster had two quad-core Intel Xeon processors running at 2.83 GHz with 4 GB of physical memory. We used 3000 seconds as the timeout interval for each invokation of BoundedSAT in UniWit, and 20 hours as the timeout interval for the overall algorithm. If an invokation of BoundedSAT in line 11 of the pseudocode timed out (after 3000 seconds), we repeated the iteration (lines 7–12 of the pseudocode of UniWit) without incrementing $i$. If the overall algorithm timed out (after 20 hours), we considered the algorithm to have failed. We used either 2 or 3 for the value of the parameter $k$ (see pseudocode of UniWit). This corresponds to restricting the pivot to few tens of witnesses for formulae with a few thousand variables. The exact values of $k$ used for a subset of the benchmarks are indicated in Table 3.1. A full analysis of the effect of parameter $k$ requires a separate study. As explained earlier, our implementation uses the family $H_{conv}(n,m,2)$ to select random hash functions in step 9 of the pseudocode.

For purposes of comparison, we also implemented and conducted experiments with algorithms BGP [18], XORSample and XORSample' [33], using CryptoMiniSAT as the SAT solver in all cases. Algorithm BGP timed out without producing any witness in all but the simplest of cases (involving less than 20 variables). This is primarily because checking whether $|R_{x,h,\alpha}| \leq 2n^2$ for a given $h \in H(n,m,n)$ and *for every* $\alpha \in \{0,1\}^m$, as required in step 10 of algorithm BGP, is computationally prohibitive for values of $n$ and $m$ exceeding few tens. Hence, we do not report any comparison with algorithm BGP. Of the algorithms XORSample and XORSample', algorithm XORSample' consistently out-performed algorithm XORSample in terms of both actual time taken and uniformity of generated witnesses. This can be largely attributed to the stringent requirement of algorithm XORSample that its input parameter $s$ must render the model count of the input formula $F$ constrained with $s$ random xor constraints to *exactly* 1. Our experiments indicated that it was extremely difficult to identify the range of values for $s$ such that it met the strict requirement of the model count being *exactly* 1. This forced us to expend significant computing resources to estimate the right value value for $s$ in almost every run, leading to huge performance overheads. Since algorithm XORSample' consistently outperformed algorithm XORSample, we focus on comparisons with only algorithm XORSample' in the subsequent discussion. Also, algorithm XORSample' has been shown to perform better than SampleSAT which in turn was shown to perform better than the algorithms based on belief networks [47].

Thus, we report results with the best of breed algorithm XORSample′. Note that our benchmarks, when viewed as Boolean circuits, had upto 695 circuit inputs, and 21 of them had more than 95 inputs each. While UniWit and XORSample′ completed execution on all these benchmarks, we could not build ROBDDs for 18 of the above 21 benchmarks within our timeout limit and with 4GB of memory. Hence no comparison with ROBDD-based techniques is reported.

## 3.5 Results

The results are presented here only for a small subset of benchmarks for lack of space. The tool and the complete set of results on over 200 benchmarks are available at http://www.cfdvs.iitb.ac.in/reports/UniWit. Table 3.1 presents results of our experiments comparing performance and uniformity of generated witnesses for UniWit and XORSample′ on a subset of benchmarks.

The first three columns in Table 3.1 give the name, number of variables and number of clauses of the benchmarks represented as CNF formulae. The columns grouped under UniWit give details of runs of UniWit, while those grouped under XORSample′ give details of runs of XORSample′. For runs of UniWit, the column labeled "$k$" gives the value of the parameter $k$ used in the corresponding experiment. The column labeled "Range ($i$)" shows the range of values of $i$ when the loop in lines 7–12 of the pseudocode (see Section 3.2) terminated in 100 independent runs of the algorithm on the benchmark under consideration. Significantly, this range is uniformly narrow for all our experiments with UniWit. As a result, leapfrogging $i$ is very effective for UniWit.

The column labeled "Run Time" under UniWit in Table 3.1 gives run times in seconds, separated as $time_1 + time_2$, where $time_1$ gives the average time (over 100 in-dependent runs) to obtain the first sample and to identify the lower bound of *i* for leapfrogging in later runs, while *time*$_2$ gives the average time to get an additional sample once we leapfrog *i*. Our experiments clearly show that leapfrogging *i* reduces run-times by almost an order of magnitude in most cases. We also report "Run Time" for XORSample′, where times are again reported as *time*$_1$ + *time*$_2$. In this case, *time*$_1$ gives the average time (over 100 independent runs) taken to find the value of the parameter *s* in algorithm XORSample′ using a binary search technique, as outlined in a footnote in [33]. As can be seen from Table 3.1, this is a computationally expensive step, and often exceeds *time*$_1$ under UniWit by more than two to three orders of magnitude. Once the range of the parameter *s* is identified from the first 100 independent runs, we use the lower bound of this range to leapfrog *s* in subsequent runs of XORSample′ on the same problem instance. The values of *time*$_2$ under "Run Time" for XORSample′ give the average time taken to generate witnesses after leapfrogging *s*. Note that the difference between *time*$_2$ values for UniWit and XORSample′ algorithms is far less pronounced than the difference between *time*$_1$ values. In addition, the *time*$_1$ values for XORSample′ are two to four orders of magnitude larger than the corresponding *time*$_2$ values, while this factor is almost always less than an order of magnitude for UniWit. Therefore, the total time taken for $n_1$ runs without leapfrogging, followed by $n_2$ runs with leapfrogging for XORSample′ far exceeds that for UniWit, even for $n_1 = 100$ and $n_2 \approx 10^6$. This illustrates the significant practical efficiency of UniWit vis-a-vis XORSample′.

| Benchmark | #var | Clauses | k | Range (i) | UniWit Average Run Time (s) | Variance | XORSample' Average Run Time (s) | Variance |
|---|---|---|---|---|---|---|---|---|
| case_3_b14 | 779 | 2480 | 2 | [34,35] | 49.29+5.27 | 1.58 | 15061.85+59.31 | 3.47 |
|  |  |  | 3 | [36,37] | 19.32+1.44 |  |  |  |
| case_2_b14 | 519 | 1607 | 3 | [38,39] | 22.13+2.09 | 0.57 | 18005.58+0.73 | 9.51 |
| case203 | 214 | 580 | 3 | [42,44] | 16.41+1.04 | 8.98 | 18006.85+2.78 | 230.5 |
| case145 | 219 | 558 | 3 | [42,44] | 19.84+1.42 | 1.62 | 18007.18+2.99 | 2.32 |
| case14 | 270 | 717 | 2 | [44,45] | 54.07+2.33 | 0.65 | 18004.8+0.9 | 28.16 |
| case61 | 289 | 773 | 3 | [44,46] | 30.39+5.49 | 1.33 | 18009.1+4.4 | 11.92 |
| case9 | 302 | 821 | 3 | [45,47] | 25.64+1.54 | 2.07 | 18004.79+0.87 | 46.15 |
| case10 | 351 | 946 | 2 | [60,61] | 204.93+17.99 | 2.68 | 18008.42+4.85 | 10.56 |
| case15 | 319 | 842 | 3 | [61,63] | 91.84+14.64 | 2.61 | 18008.34+5.08 | 11.04 |
| case140 | 488 | 1222 | 3 | [99,101] | 288.63+23.53 | 1.41 | 21214.85+200.64 | 6.71 |
| squaring14 | 5397 | 18141 | 3 | [28,30] | 2399.19+1243.81 |  | 7089.6+2088.46 |  |
| squaring7 | 5567 | 18969 | 3 | [26,29] | 2358.45+1720.49 |  | 4841.4+2340.84 |  |
| case39 | 590 | 1789 | 2 | [50,50] | 710.65+85.22 |  | 18159.12+138.22 |  |
| case_2_ptb | 7621 | 24889 | 3 | [72,73] | 1643.2+225.41 |  | 22251.8+177.61 |  |
| case_1_ptb | 7624 | 24897 | 2 | [70,70] | 17295.45+454.64 |  | 22346.64+204.07 |  |
|  |  |  | 3 | [72,73] | 1639.16+219.87 |  |  |  |

Table 3.1 : Performance comparison of UniWit and XORSample'

Table 3.1 also reports the scaled statistical variance of relative frequencies of witnesses generated by $5 \times 10^4$ runs of the two algorithms on several benchmarks. The scaled statistical variance is computed as $\frac{K}{N-1} \sum_{i=1}^{N} \left( f_i - \left( \frac{\sum_{i=1}^{N} f_i}{N} \right) \right)^2$, where *N* denotes the number of distinct witnesses generated, $f_i$ denotes the relative frequency of the *i*$^{th}$ witness, and $K$ ($10^{10}$) denotes a scaling constant used to facilitate easier comparison. The smaller the scaled variance, the more uniform is the generated distribution. Unfortunately, getting a reliable estimate of the variance requires generating witnesses from runs that sample the witness space sufficiently well. While we could do this for several benchmarks (listed towards the top of Table 3.1), other benchmarks (listed towards the bottom of Table 3.1) had too large witness spaces to conduct these experiments within available resources. For those benchmarks where we have variance data, we observe that the variance obtained using XORSample′ is larger (by upto a factor of 43) than those obtained using UniWit in almost all cases. Overall, our experiments indicate that UniWit always works significantly faster and gives more (or comparably) uniformly distributed witnesses vis-a-vis XORSample′ in almost all cases. We also measured the probability of success of UniWit for each benchmark as the ratio of the number of runs for which the algorithm did not return ⊥ to the total number of runs. We found that this exceeded 0.6 for every benchmark using UniWit which is significantly higher than the conservative bounds presented in 3.3

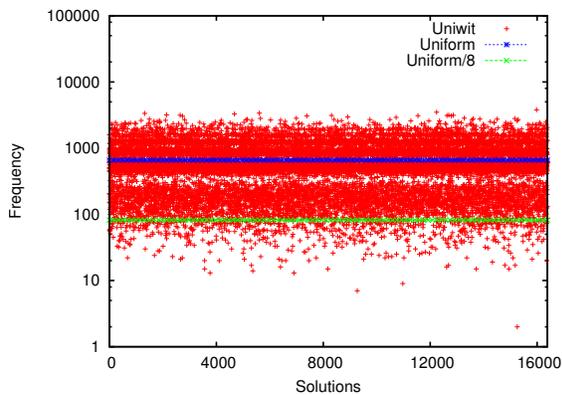 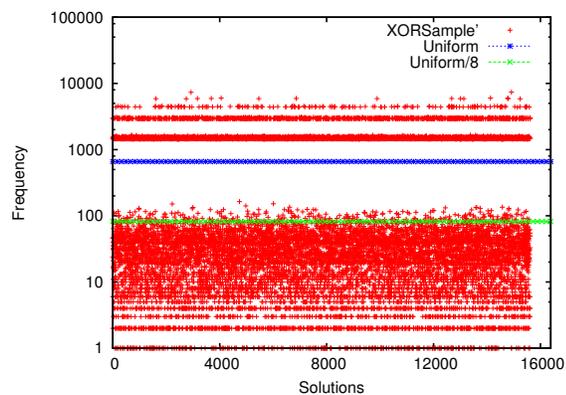

Figure 3.1 : Sampling by UniWit (k=2)    Figure 3.2 : Sampling by XORSample′

As an illustration of the difference in uniformity of witnesses generated by UniWit and XORSample′, Figures 3.1 and 3.2 depict the frequencies of appearance of various witnesses using these two algorithms for an input CNF formula (case110) with 287 variables and 16,384 satisfying assignments. The horizontal axis in each figure represents witnesses numbered suitably, while the vertical axis represents the generated frequencies of witnesses. The frequencies were obtained from $10.8 \times 10^6$ successful runs of each algorithm. The clustering of points in Figures 3.1 and 3.2 visually depict the differences in uniformity of witnesses generated by the two algorithms. Interestingly, XORSample′ could find only 15,612 solutions (note the empty vertical band at the right end of Figure 3.2), while UniWit found all 16,384 solutions. Further, XORSample′ generated each of 15 solutions more than 5,500 times, and more than 250 solutions were generated only once. No such major deviations from uniformity were however observed in the frequencies generated by UniWit. We also found that 15624 out of 16384 (i.e. 95.36%) witnesses generated by UniWit had frequencies in excess of $N_{unif}/8$, where $N_{unif} = 10.8 \times 10^6/16384 \approx 659$. In contrast, only 6047 (i.e. 36.91%) witnesses generated by XORSample′ had frequencies in excess of $N_{unif}/8$.

## 3.6 Conclusion

We described UniWit, an algorithm that near-uniformly samples random witnesses of Boolean formulas. We showed that the algorithm scales to reasonably large problems. Although we focused on SAT formulas, the core ideas introduced in this chapter are quite general and can be extended to systems with richer set of constraints like SMT constraints or quantified formulas. In the next chapter, we show how the basic ideas introduced in this Chapter can be extended to solve another important problem: that of approximately counting the number of solutions for CNF SAT formulas.

# Chapter 4

# Model Counting

Propositional Model counting, also known as #SAT, concerns counting the number of models (satisfying truth assignments) of a given propositional formula. This is an important problem with applications to a wide variety of domains ranging from probabilistic reasoning to multi-agent adversarial planning [12, 19, 13, 20]. Since a large body of problems arising from various domains (such as probabilistic reasoning) are encoded as CNF SAT constraints, we focus on the problem of model counting for CNF SAT formulas. The technique developed in in this chapter follows the limited-independence hashing based approach introduced earlier in this thesis.

#SAT has been the subject of extensive theoretical investigation since its introduction by Valiant [5]. Theoretical investigations of #SAT have led to the discovery of deep connections in complexity theory [6, 7, 8]. Simon [7] showed #SAT to be #P-complete, where #P is the set of counting problems associated with decision problems in the complexity class $\mathscr{NP}$. Subsequently, Angluin [6] observed the equality of the complexity classes $\mathsf{P}^{\mathsf{PP}}$ and $\mathsf{P}^{\#\mathsf{P}}$, where $\mathsf{P}^{\mathscr{C}}$ denotes the class of decision problems solvable in polynomial-time with access to an oracle for queries in $\mathscr{C}$ and PP denotes the class of decision problems that can be solved in polynomial time by a probabilistic Turing machine with an error probability of less than $1/2$. Building on this equality, Toda [8] showed that $\mathsf{P}^{\#\mathsf{P}}$ is as hard as the polynomial hierarchy; in fact, the polynomial-time machine only needs to make one #SAT query to solve any problem in the polynomial hierarchy. A consequence of Toda's result is that decid-

ing a quantified Boolean formula with a constant number of ∃ and ∀ quantification alternations can be reduced to #SAT. Thus, there is strong evidence for the hardness of #SAT.

On the implementation front, the earliest approaches to #SAT were based on Davis-Putnam-Loveland-Longemann(or DPLL)-style SAT solvers and computed exact counts. These approaches consisted of incrementally counting the number of solutions by adding appropriate multiplication factors after a partial solution was found. This idea was formalized by Birnbaum and Lozinkii [55] in their model counter CDP. Subsequent model counters such as Relsat [56], Cachet [57] and sharpSAT [58] improved upon this idea by using several optimizations such as component caching, clause learning, look-ahead and the like. Techniques based on Boolean Decision Diagrams and their variants [59, 60], or d-DNNF formulae [61], have also been used to compute exact model counts. Although exact model counters have been successfully used in small to medium-sized problems, scaling to larger problem instances has posed significant challenges in practice. Consequently, a large class of practical applications has remained beyond the reach of exact model counters.

In many applications of model counting, such as in probabilistic reasoning, the exact model count may not be critically important, and approximate counts are sufficient. Even when exact model counts are important, the inherent complexity of the problem may force one to work with approximate counters in practice. Thus, designing approximate counters that scale to practical problem sizes is an important problem. Earlier work on approximate counters has been restricted largely to theoretical treatments of the problem [23, 14]. The only counter in this category that we are aware of as having been implemented is due to Karp and Luby [62]. Karp and Luby presented a fully polynomial randomized approximation scheme for counting models of a DNF, *but not CNF* formulas.

The counting problems for both CNF and DNF formulae are #P-complete. While the DNF representation suits some applications, most modern applications of model counting (e.g. probabilistic inference) use the CNF representation. Although *exact* counting for DNF and CNF formulae are polynomially inter-reducible, there is no known polynomial reduction for the corresponding *approximate*-counting problems. In fact, Karp and Luby remark in [21] that it is highly unlikely that their randomized approximate algorithm for DNF formulae can be adapted to work for CNF formulas. Thus, there has been no prior implementation of approximate model counters for CNF formulae *that scales in practice*. In this chapter, we present the first such counter. The primary focus of this chapter is on $(\varepsilon, \delta)$ counters for #SAT. As in [14], our algorithm runs in random polynomial time using an oracle for SAT. Our extensive experiments show that our algorithm scales, with low error, to formulas arising from several application domains involving tens of thousands of variables.

The remainder of this chapter is as follows. We present related work in Section 4.1. In Section 4.2, we present our algorithm, followed by its analysis in Section 4.3. Section 4.4 discusses our experimental methodology, followed by experimental results in Section 4.5. We finally conclude in Section 4.6 with a discussion on the extension of the techniques presented in this chapter to other problems in #P class.

## 4.1 Prior Work

In this section, we briefly review various works proposed in the literature for model counting in SAT. Finally we discuss in detail one algorithm that comes closest to our work.

**Guarantee-less and Bounding Model Counters:** Several approaches have been proposed over the years to counter the scalability challenge posed by model counting. Researchers have proposed guarantee-less counters which do not offer guarantees at all but they can be very efficient and provide good approximations in practice. Examples of guarantee-less counters include ApproxCount [40], SearchTreeSampler [63], SE [64] and SampleSearch [65]. The large majority of approximate counters used in practice are bounding counters. Notable examples include SampleCount [66], BPCount [27], MBound (and Hybrid-MBound) [30], and MiniCount [27]. As noted in Section 2.1, these upper bounding counters fail to provide any tolerance guarantees.

**Approximate Model Counters:** Bounding both the tolerance and confidence of approximate model counts is extremely valuable in applications like probabilistic inference. In [23], Stockmeyer showed that counting models within a specified tolerance factor can be achieved in deterministic polynomial time using a $\Sigma_2^p$-oracle. Building on Stockmeyer's result, Jerrum, Valiant and Vazirani [14] showed that counting models of CNF formulas within a specified tolerance factor can be solved in random polynomial time using an oracle for SAT. Earlier work on approximate counters has been restricted largely to theoretical treatments of the problem. The only counter in this category that we are aware of as having been implemented is due to Karp and Luby [62]. Karp and Luby's original implementation was designed to estimate reliabilities of networks with failure-prone links. However, the underlying Monte Carlo engine can be used to approximately count models of DNF, *but not CNF*, formulas.

**Hashing-based approaches:** Bounding counters and guarantee-less counters that have been implemented and applied to practical problem instances in the past have overwhelmingly used Monte Carlo techniques. In contrast, our algorithm uses a hashing-based approach, originally pioneered by Sipser in [34], and subsequently been used in theoretical [67, 18] and practical [33, 24, 30] treatments of approximate-counting and (near-)uniform sampling. Earlier implementations of counters that use the hashing-based approach are MBound and Hybrid-MBound [30]. Both these counters use the same family of hashing functions, i.e., $H_{xor}(n,m,3)$, that we use. Nevertheless, there are significant differences between our algorithm and those of MBound and Hybrid-MBound. Specifically, we are able to exploit properties of the $H_{xor}(n,m,3)$ family of hash functions to obtain a fully polynomial $(\varepsilon,\delta)$-counter with respect to a SAT oracle. In contrast, both MBound and Hybrid-MBound are bounding counters, and cannot provide bounds on tolerance. In addition, our algorithm requires no additional parameters beyond the tolerance $\varepsilon$ and confidence $1-\delta$. In contrast, the performance and quality of results of both MBound and Hybrid-MBound, depend crucially on some hard-to-estimate parameters. It has been our experience that the right choice of these parameters is often domain dependent and difficult.

### 4.1.1  JVV Algorithm

We now discuss an algorithm that comes closest to our work. Jerrum, Valiant and Vazirani [14] showed that if $R$ is a self-reducible $\mathcal{NP}$ relation (such as SAT), the problem of generating models *almost uniformly* is polynomially inter-reducible with approximately counting models. Recall from Section 2.1.1 that almost-uniform generation requires that if $x$ is a problem instance, then for every $y \in R_x$, we have $(1+\varepsilon)^{-1}\varphi(x) \le \Pr[y \text{ is generated}] \le (1+\varepsilon)\varphi(x)$, where $\varepsilon > 0$ is the specified tolerance and $\varphi(x)$ is an appropriate function. Given an almost uniform generator $\mathcal{G}$ for $R$, an input $x$, a tolerance bound $\varepsilon$ and a confidence bound $1-\delta$, it is shown in [14] that one can obtain an $(\varepsilon,\delta)$-counter for $R$ by invoking $\mathcal{G}$ polynomially (in $|x|$, $1/\varepsilon$ and $\log_2(1/\delta)$) many times, and by using the generated samples to estimate $|R_x|$. For convenience of exposition, we refer to this approximate-counting algorithm as the JVV algorithm (after the last names of the authors).

An important feature of the JVV algorithm is that it uses the almost-uniform generator $\mathscr{G}$ as a black box. Specifically, the details of how $\mathscr{G}$ works is of no consequence. Prima facie, this gives us freedom in the choice of $\mathscr{G}$ when implementing the JVV algorithm. Unfortunately, while there are theoretical constructions of uniform generators in [18], we are not aware of any implementation of an almost-uniform generator that scales to CNF formulas involving thousands of variables. The authors of [14] give a theoretical description of how an almost-uniform generator for a self-reducible $\mathscr{NP}$ relation $R$ can be obtained from an $(\varepsilon, \delta)$-counter for $R$. Unfortunately, this implies a cyclic dependency on the existence of an $(\varepsilon, \delta)$ counter. The lack of a scalable and almost-uniform generator presents a significant hurdle in implementing the JVV algorithm for practical applications. It is worth asking if we can make the JVV algorithm work without requiring $\mathscr{G}$ to be an almost-uniform generator. A closer look at the proof of correctness of the JVV algorithm [14] shows that it crucially relies on the ability of $\mathscr{G}$ to ensure that the probabilities of generation of any two distinct models of $x$ differ by a factor in $O(\varepsilon^2)$. As discussed in the previous chapter, existing algorithms for randomly generating models either provide this guarantee but scale very poorly in practice (e.g., the algorithms in [18, 11]), or scale well in practice without providing the above guarantee (e.g., the algorithms in [24, 33, 35]). The near-uniform generator UniWit, proposed in the previous chapter, provides only a near-uniformity guarantee. Therefore, using an existing generator as a black box in the JVV algorithm would not give us an $(\varepsilon, \delta)$ model counter that scales in practice. The primary contribution of this chapter is to show that a scalable $(\varepsilon, \delta)$-counter can indeed be designed by using the same insights that went into the design of the *near uniform* generator UniWit, but without using the generator as a black box in the approximate-counting algorithm.

Note that near uniformity, as defined in Section 2.1.1, is a more relaxed notion of uniformity than almost uniformity. We leave the question of whether a near-uniform generator can be used as a black box to design an $(\varepsilon, \delta)$ counter as part of future work.

The central idea of UniWit, which is also shared by our approximate model counter, is the use of $r$-wise independent hashing functions to randomly partition the space of all models of a given problem instance into "small" cells. This idea was first proposed in [18], but there are three novel insights that allow UniWit to scale better than other hashing-based sampling algorithms [18, 33], while still providing guarantees on the quality of sampling. These insights are: (i) the use of computationally efficient linear hashing functions with low degrees of independence, (ii) requirement of only a randomly chosen to be "small" instead of *every* cell being "small" and (iii) a drastic reduction in the size of "small" cells, from $n^2$ in [18] to $n^{1/k}$ (for $2 \leq k \leq 3$) in UniWit (where $n$ is the number of propositional variables), and even further to a constant in the this chapter. We continue to use these key insights in the design of our approximate model counter, although UniWit is not used explicitly in the model counter.

## 4.2 The ApproxMC Algorithm

We now describe our approximate model counting algorithm, called ApproxMC. We use pairwise independent linear hashing functions from the $H_{xor}(n,m,3)$ family, for an appropriate $m$, to randomly partition the set of models of an input formula into "small" cells. The choice of $H_{xor}(n,m,3)$ is arbitrary and any other pairwise independent hashing family such as $H_{conv}(n,m,2)$ also suffices. In order to test whether the generated cells are indeed small, we choose a random cell and check if it is non-empty and has no more than *pivot* elements, where *pivot* is a threshold that depends only on the tolerance bound $\varepsilon$. If the chosen cell is not small, we randomly partition the set of models into twice as many cells as before by choosing a random hashing function from the family $H_{xor}(n, m+1, 3)$. The above procedure is repeated until either a randomly chosen cell is found to be non-empty and small, or the number of cells exceeds $\frac{2^{n+1}}{pivot}$. If all cells that were randomly chosen during the above process were either empty or not small, we report a counting failure and return $\bot$. Otherwise, the size of the cell last chosen is multiplied by the number of cells to obtain an $\varepsilon$-approximate estimate of the model count.

The procedure outlined above forms the core engine of ApproxMC. For convenience of exposition, we implement this core engine as a function ApproxMCCore. The overall ApproxMC algorithm simply invokes ApproxMCCore sufficiently many times, and returns the median of the non-$\bot$ values returned by ApproxMCCore. The pseudo-code for algorithm ApproxMC is shown below.

Algorithm ApproxMC takes as inputs a CNF formula $F$, a tolerance $\varepsilon$ ($0 < \varepsilon \le 1$) and $\delta$ ($0 < \delta \le 1$) such that the desired confidence is $1 - \delta$. It computes two key parameters: (i) a threshold *pivot* that depends only on $\varepsilon$ and is used in ApproxMCCore to determine the size of a "small" cell, and (ii) a parameter $t$ ($\ge 1$) that depends only on $\delta$ and is used to determine the number of times ApproxMCCore is invoked. The particular choice of functions to compute the parameters *pivot* and $t$ aids us in proving theoretical guarantees for ApproxMC in Section 4.3. Note that *pivot* is in $\mathcal{O}(1/\varepsilon^2)$ and $t$ is in $\mathcal{O}(\log_2(1/\delta))$. All non-$\bot$ estimates of the model count returned by ApproxMCCore are stored in the list $C$. The function AddToList($C, c$) updates the list $C$ by adding the element $c$. The final estimate of the model count returned by ApproxMC is the median of the estimates stored in $C$, computed using FindMedian($C$).

---

**Algorithm 4** ApproxMC($F, \varepsilon, \delta$)
1: *counter* $\leftarrow$ 0; $C \leftarrow$ emptyList;
2: *pivot* $\leftarrow$ 2 $\times$ ComputeThreshold($\varepsilon$);
3: $t \leftarrow$ ComputeIterCount($\delta$);
4: **repeat**
5:    $c \leftarrow$ ApproxMCCore($F, pivot$);
6:    *counter* $\leftarrow$ *counter* $+$ 1;
7:    **if** ($c \neq \bot$) **then**
8:       AddToList($C, c$);
9: **until** (*counter* $> t$)
10: *finalCount* $\leftarrow$ FindMedian($C$);
11: **return** *finalCount*;

---

**Algorithm 5** ComputeThreshold($\varepsilon$)
1: **return** $\left\lceil 3e^{1/2} \left(1 + \frac{1}{\varepsilon}\right)^2 \right\rceil$;

---

**Algorithm 6** ComputeIterCount($\delta$)
   **return** $\lceil 35 \log_2(3/\delta) \rceil$;

We assume that if the list $C$ is empty, FindMedian($C$) returns $\bot$.

The pseudocode for ApproxMCCore is presented in Algorithm 7. Algorithm ApproxMCCore

---

**Algorithm 7** ApproxMCCore($F, pivot$)

    /* Assume $z_1, \ldots z_n$ are the variables of $F$ */

1:  $S \leftarrow$ BoundedSAT($F, pivot + 1$);

2:  **if** ( **then**$|S| \leq pivot$) **return** $|S|$;

3:  **else**

4:     $l \leftarrow \lfloor \log_2(pivot) \rfloor - 1$; $i \leftarrow l - 1$;

5:     **repeat**

6:         $i \leftarrow i + 1$;

7:         Choose $h$ at random from $H_{xor}(n, i - l, 3)$;

8:         Choose $\alpha$ at random from $\{0, 1\}^{i-l}$;

9:         $S \leftarrow$ BoundedSAT($F \wedge (h(z_1, \ldots z_n) = \alpha), pivot + 1$);

10:    **until** $(1 \leq |S| \leq pivot)$ or $(i = n)$;

11:    **if** $(|S| > pivot$ **or** $|S| = 0)$ **then return** $\bot$ ;

12:    **else return** $|S| \cdot 2^{i-l}$;

---

takes as inputs a CNF formula $F$ and a threshold *pivot*, and returns an approximate estimate of the model count of $F$. We assume that ApproxMCCore has access to a function BoundedSAT that takes as inputs a proposition formula $F'$ that is the conjunction of a CNF formula and xor constraints, as well as a threshold $v \geq 0$. BoundedSAT($F', v$) returns a set $S$ of models of $F'$ such that $|S| = \min(v, \#F')$. If the model count of $F$ is no larger than *pivot*, then ApproxMCCore returns the exact model count of $F$ in line 3 of the pseudo-code. Otherwise, it partitions the space of all models of $F$ using random hashing functions from $H_{xor}(n, i - l, 3)$ and checks if a randomly chosen cell is non-empty and has at most *pivot* elements. Lines 8–10 of the repeat-until loop in the pseudo-code implement this functionality. The loop terminates if either a randomly chosen cell is found to be small and non-empty, or if the number of cells generated exceeds $\frac{2^{n+1}}{pivot}$ (i.e. $i = n$ in line 11). In all cases, unless the cell that was chosen last is empty or not small, we multiply its size by the number of cells generated by the corresponding hashing function to compute an estimate of the model count. If, however, all randomly chosen cells turn out to be empty or not small, we report a counting error by returning $\perp$.

**Implementation issues:** As in algorithm UniWit, there are two steps in algorithm ApproxMCCore (lines 8 and 9 of the pseudocode) where random choices are made. Recall from Section 2.3 that choosing a random hash function from $H_{xor}(n,m,3)$ requires choosing random bit-vectors. It is straightforward to implement these choices and also the choice of a random $\alpha \in \{0,1\}^{i-l}$ in line 9 of the pseudo-code, if we have access to a source of independent and uniformly distributed random bits. Our implementation uses pseudo-random sequences of bits generated from nuclear decay processes and made available at HotBits [51]. We download and store a sufficiently long sequence of random bits in a file, and access an appropriate number of bits sequentially whenever needed. We defer experimenting with sequences of bits obtained from pseudo-random generators to a future study.

In lines 1 and 10 of the pseudo-code for algorithm ApproxMCCore, we invoke the function BoundedSAT. Note that if $h$ is chosen randomly from $H_{xor}(n,m,3)$, the formula for which we seek models is the conjunction of the original (CNF) formula and xor constraints encoding the inclusion of each witness in $h^{-1}(\alpha)$. We therefore use a SAT solver optimized for conjunctions of xor constraints and CNF clauses as the back-end engine. Specifically, we use CryptoMiniSAT (version 2.9.2) [50], which also allows passing a parameter indicating the maximum number of witnesses to be generated.

Recall that ApproxMCCore is invoked $t$ times with the same arguments in algorithm ApproxMC. Repeating the loop of lines 6–11 in the pseudocode of ApproxMCCore in each invocation can be time consuming if the values of $i-l$ for which the loop terminates are large. In the previous chapter, a heuristic called *leap-frogging* was proposed to overcome this bottleneck in practice. With leap-frogging, we register the smallest value of $i-l$ for which the loop terminates during the first few invocations of ApproxMCCore. In all subsequent invocations of ApproxMCCore with the same arguments, we start iterating the loop of lines 6–11 by initializing $i-l$ to the smallest value registered from earlier invocations. Our experiments indicate that leap-frogging is extremely efficient in practice and leads to significant savings in time after the first few invocations of ApproxMCCore.

## 4.3 Analysis of ApproxMC

We now formally prove that algorithm ApproxMC is an $(\varepsilon, \delta)$ counter for $0 < \varepsilon \leq 1$ and $0 < \delta \leq 1$. Subsequently we also present a complexity analysis of ApproxMC.

Let $F$ be a CNF propositional formula with $n$ variables. The next two lemmas show that algorithm ApproxMCCore, when invoked from ApproxMC with arguments $F$, $\varepsilon$ and $\delta$, behaves like an $(\varepsilon, d)$ model counter for $F$, for a fixed confidence $1-d$ (possibly different from $1-\delta$). Throughout this section, we use the notations $R_F$ and $R_{F,h,\alpha}$ introduced in Section 2.1.

### 4.3.1 Approximation Guarantees

**Sketch of Analysis**

Theorem 4.3.4 shows that ApproxMC is $(\varepsilon, \delta)$-counter. The proof of Theorem 4.3.4 depends on Lemma 4.3.1 and Lemma 4.3.2. Lemma 4.3.1 shows that the a non-$\perp$ count returned by ApproxMCCore lies within desired tolerance but with confidence of only $1 - e^{-3/2}$. Lemma 4.3.2 provides an upper $(1 - e^{-3/2})$ bound on the event that $\perp$ is returned by ApproxMCCore. Combining Lemmas 4.3.2 and 4.3.1 yields Theorem 4.3.3, which states that count returned by ApproxMCCore lines within desired tolerance with confidence at least $(1 - e^{-3/2})^2$. Finally in the proof of Theorem 4.3.4 shows that making $t$ calls to ApproxMCCore and the act of choosing median allows us to achieve the desired confidence. Furthermore, Theorem 4.3.5 provides the time complexity for ApproxMC.

**Lemma 4.3.1.** *Let algorithm* ApproxMCCore, *when invoked from* ApproxMC, *return $c$ with $i$ being the final value of the loop counter in* ApproxMCCore. *Then,* $\Pr\left[(1+\varepsilon)^{-1} \cdot |R_F| \leq c \leq (1+\varepsilon) \cdot |R_F| \,\middle|\, c \neq \perp \text{ and } i \leq \log_2 |R_F|\right] \geq 1 - e^{-3/2}.$

*Proof.* Referring to the pseudocode of ApproxMCCore, the lemma is trivially satisfied if $|R_F| \leq pivot$. Therefore, the only non-trivial case to consider is when $|R_F| > pivot$ and ApproxMCCore returns from line 13 of the pseudo-code. In this case, the count returned is $2^{i-l} \cdot |R_{F,h,\alpha}|$, where $l = \lfloor \log_2(pivot) \rfloor - 1$ and $\alpha, i$ and $h$ denote (with abuse of notation) the values of the corresponding variables and hash functions in the final iteration of the repeat-until loop in lines 6–11 of the pseudo-code.

For simplicity of exposition, we assume henceforth that $\log_2(pivot)$ is an integer. A more careful analysis removes this restriction with only a constant factor scaling of the probabilities. From the pseudo-code of ApproxMCCore, we know that $pivot =$

$2\left\lceil 3e^{1/2}\left\lceil 1+\frac{1}{\varepsilon}\right\rceil^2\right\rceil$.

Furthermore, the value of $i$ is always in $\{l,\ldots n\}$. Since $pivot < |R_F| \leq 2^n$ and $l = \lfloor \log_2 pivot \rfloor - 1$, we have $l < \log_2 |R_F| \leq n$. The lemma is now proved by showing that for every $i$ in $\{l,\ldots \lfloor \log_2 |R_F|\rfloor\}$, $h \in H(n, i-l, 3)$ and $\alpha \in \{0,1\}^{i-l}$, we have $\Pr\left[(1+\varepsilon)^{-1} \cdot |R_F| \leq 2^{i-l}|R_{F,h,\alpha}| \leq (1+\varepsilon) \cdot |R_F|\right] \geq (1 - e^{-3/2})$.

For every $y \in \{0,1\}^n$ and for every $\alpha \in \{0,1\}^{i-l}$, define an indicator variable $\gamma_{y,\alpha}$ as follows: $\gamma_{y,\alpha} = 1$ if $h(y) = \alpha$, and $\gamma_{y,\alpha} = 0$ otherwise. Let us fix $\alpha$ and $y$ and choose $h$ uniformly at random from $H(n, i-l, 3)$. The random choice of $h$ induces a probability distribution on $\gamma_{y,\alpha}$, such that $\Pr[\gamma_{y,\alpha} = 1] = \Pr[h(y) = \alpha] = 2^{-(i-l)}$, and $\mathsf{E}[\gamma_{y,\alpha}] = \Pr[\gamma_{y,\alpha} = 1] = 2^{-(i-l)}$. In addition, the pairwise independence of hash functions chosen from $H(n, i-l, 3)$ implies that for every distinct $y_a$, and $y_b \in R_F$, the random variables $\gamma_{y_a,\alpha}$, and $\gamma_{y_b,\alpha}$ are pairwise independent.

Let $\Gamma_\alpha = \sum_{y \in R_F} \gamma_{y,\alpha}$ and $\mu_\alpha = \mathsf{E}[\Gamma_\alpha]$. Clearly, $\Gamma_\alpha = |R_{F,h,\alpha}|$ and $\mu_\alpha = \sum_{y \in R_F} \mathsf{E}[\gamma_{y,\alpha}] = 2^{-(i-l)}|R_F|$. Since $|R_F| > pivot$ and $i \leq \log_2 |R_F|$, using the expression for $pivot$, we get $3 \leq \left\lfloor e^{-1/2}(1+\frac{1}{\varepsilon})^{-2} \cdot \frac{|R_F|}{2^{i-l}}\right\rfloor$. Therefore, using Theorem 2.2.1, $\Pr\left[|R_F| \cdot \left(1 - \frac{\varepsilon}{1+\varepsilon}\right) \leq 2^{i-l}|R_{F,h,\alpha}| \leq (1+\frac{\varepsilon}{1+\varepsilon})|R_F|\right] \geq 1 - e^{-3/2}$. Simplifying and noting that $\frac{\varepsilon}{1+\varepsilon} < \varepsilon$ for all $\varepsilon > 0$, we obtain $\Pr\left[(1+\varepsilon)^{-1} \cdot |R_F| \leq 2^{i-l}|R_{F,h,\alpha}| \leq (1+\varepsilon) \cdot |R_F|\right] \geq 1 - e^{-3/2}$. □

**Lemma 4.3.2.** *Given $|R_F| > pivot$, the probability that an invocation of* ApproxMCCore *from* ApproxMC *returns non-$\perp$ with $i \leq \log_2 |R_F|$, is at least $1 - e^{-3/2}$.*

*Proof.* Let us denote $\log_2 |R_F| - l = \log_2 |R_F| - (\lfloor \log_2(pivot) \rfloor - 1)$ by $m$. Since $|R_F| > pivot$ and $|R_F| \leq 2^n$, we have $l < m + l \leq n$. Let $p_i$ ($l \leq i \leq n$) denote the conditional probability that ApproxMCCore($F, pivot$) terminates in iteration $i$ of the repeat-until loop (lines 6–11 of the pseudo-code) with $1 \leq |R_{F,h,\alpha}| \leq pivot$, given $|R_F| > pivot$. Since the choice of $h$ and $\alpha$ in each iteration of the loop are independent of those in

previous iterations, the conditional probability that ApproxMCCore($F$,$pivot$) returns non-$\perp$ with $i \leq \log_2 |R_F| = m+l$, given $|R_F| > pivot$, is $p_l + (1-p_l)p_{l+1} + \cdots + (1-p_l)(1-p_{l+1})\cdots(1-p_{m+l-1})p_{m+l}$. Let us denote this sum by $P$. Thus, $P = p_l + \sum_{i=l+1}^{m+l}\prod_{k=l}^{i-1}(1-p_k)p_i \geq \left(p_l + \sum_{i=l+1}^{m+l-1}\prod_{k=l}^{i-1}(1-p_k)p_i\right)p_{m+l} + \prod_{s=l}^{m+l-1}(1-p_s)p_{m+l} = p_{m+l}$. The lemma is now proved by using Theorem 2.2.1 to show that $p_{m+l} \geq 1 - e^{-3/2}$.

It was shown in Lemma 2.2.1 that $\Pr\left[(1+\varepsilon)^{-1} \cdot |R_F| \leq 2^{i-l}|R_{F,h,\alpha}| \leq (1+\varepsilon) \cdot |R_F|\right] \geq 1 - e^{-3/2}$ for every $i \in \{l,\ldots\lfloor\log_2|R_F|\rfloor\}$, $h \in H(n, i-l, 3)$ and $\alpha \in \{0,1\}^{i-l}$. Substituting $\log_2|R_F| = m+l$ for $i$, re-arranging terms and noting that the definition of $m$ implies $2^{-m}|R_F| = pivot/2$, we get $\Pr\left[(1+\varepsilon)^{-1}(pivot/2) \leq |R_{F,h,\alpha}| \leq (1+\varepsilon)(pivot/2)\right] \geq 1 - e^{-3/2}$. Since $0 < \varepsilon \leq 1$ and $pivot > 4$, it follows that $\Pr\left[1 \leq |R_{F,h,\alpha}| \leq pivot\right] \geq 1 - e^{-3/2}$. Hence, $p_{m+l} \geq 1 - e^{-3/2}$. □

**Theorem 4.3.3.** *Let an invocation of* ApproxMCCore *from* ApproxMC *return $c$. Then* $\Pr\left[c \neq \perp \text{ and } (1+\varepsilon)^{-1} \cdot |R_F| \leq c \leq (1+\varepsilon) \cdot |R_F|\right] \geq (1-e^{-3/2})^2 > 0.6$.

*Proof.* It is easy to see that the required probability is at least as large as $\Pr\left[c \neq \perp \text{ and } i \leq \log_2|R_F|\right]$. From Lemmas 4.3.1 and 4.3.2, the latter probability is $\geq (1-e^{-3/2})^2$.

We now turn to proving that the confidence can be raised to at least $1-\delta$ for $\delta \in (0,1]$ by invoking ApproxMCCore $\mathcal{O}(\log_2(1/\delta))$ times, and by using the median of the non-$\perp$ counts thus returned. For convenience of exposition, we use $\eta(t,m,p)$ in the following discussion to denote the probability of at least $m$ heads in $t$ independent tosses of a biased coin with $\Pr[heads] = p$. Clearly, $\eta(t,m,p) = \sum_{k=m}^{t}\binom{t}{k}p^k(1-p)^{t-k}$.

**Theorem 4.3.4.** *Given a propositional formula $F$ and parameters $\varepsilon$ ($0 < \varepsilon \leq 1$) and $\delta$ ($0 < \delta \leq 1$), suppose* ApproxMC($F,\varepsilon,\delta$) *returns $c$. Then* $\Pr\left[(1+\varepsilon)^{-1} \cdot |R_F| \leq c \leq (1+\varepsilon) \cdot |R_F|\right] \geq 1 - \delta$.

*Proof.* Throughout this proof, we assume that ApproxMCCore is invoked $t$ times from ApproxMC, where $t = \lceil 35\log_2(3/\delta) \rceil$ (see pseudocode for ComputeIterCount in Section 4.2). Referring to the pseudo-code of ApproxMC, the final count returned by ApproxMC is the median of non-$\bot$ counts obtained from the $t$ invocations of ApproxMCCore. Let *Err* denote the event that the median is not in $[(1+\varepsilon)^{-1} \cdot |R_F|, (1+\varepsilon) \cdot |R_F|]$. Let "#$non\bot = q$" denote the event that $q$ (out of $t$) values returned by ApproxMCCore are non-$\bot$. Then, $\Pr[Err] = \sum_{q=0}^{t} \Pr[Err \mid \#non\bot = q] \cdot \Pr[\#non\bot = q]$.

In order to obtain $\Pr[Err \mid \#non\bot = q]$, we define a 0-1 random variable $Z_i$, for $1 \leq i \leq t$, as follows. If the $i^{th}$ invocation of ApproxMCCore returns $c$, and if $c$ is either $\bot$ or a non-$\bot$ value that does not lie in the interval $[(1+\varepsilon)^{-1} \cdot |R_F|, (1+\varepsilon) \cdot |R_F|]$, we set $Z_i$ to 1; otherwise, we set it to 0. From Theorem 4.3.3, $\Pr[Z_i = 1] = p < 0.4$. If $Z$ denotes $\sum_{i=1}^{t} Z_i$, a necessary (but not sufficient) condition for event *Err* to occur, given that $q$ non-$\bot$s were returned by ApproxMCCore, is $Z \geq (t - q + \lceil q/2 \rceil)$. To see why this is so, note that $t - q$ invocations of ApproxMCCore must return $\bot$. In addition, at least $\lceil q/2 \rceil$ of the remaining $q$ invocations must return values outside the desired interval. To simplify the exposition, let $q$ be an even integer. A more careful analysis removes this restriction and results in an additional constant scaling factor for $\Pr[Err]$. With our simplifying assumption, $\Pr[Err \mid \#non\bot = q] \leq \Pr[Z \geq (t - q + q/2)] = \eta(t, t - q/2, p)$. Since $\eta(t, m, p)$ is a decreasing function of $m$ and since $q/2 \leq t - q/2 \leq t$, we have $\Pr[Err \mid \#non\bot = q] \leq \eta(t, t/2, p)$. If $p < 1/2$, it is easy to verify that $\eta(t, t/2, p)$ is an increasing function of $p$. In our case, $p < 0.4$; hence, $\Pr[Err \mid \#non\bot = q] \leq \eta(t, t/2, 0.4)$.

It follows from above that $\Pr[Err] = \sum_{q=0}^{t} \Pr[Err \mid \#non\bot = q] \cdot \Pr[\#non\bot = q] \leq \eta(t, t/2, 0.4) \cdot \sum_{q=0}^{t} \Pr[\#non\bot = q] = \eta(t, t/2, 0.4)$. Since $\binom{t}{t/2} \geq \binom{t}{k}$ for all $t/2 \leq k \leq t$, and since $\binom{t}{t/2} \leq 2^t$, we have $\eta(t, t/2, 0.4) = \sum_{k=t/2}^{t} \binom{t}{k} (0.4)^k (0.6)^{t-k} \leq \binom{t}{t/2} \sum_{k=t/2}^{t} (0.4)^k (0.6)^{t-k}$ $\leq 2^t \sum_{k=t/2}^{t}(0.6)^t(0.4/0.6)^k \leq 2^t \cdot 3 \cdot (0.6 \times 0.4)^{t/2} \leq 3 \cdot (0.98)^t$. Since $t = \lceil 35\log_2(3/\delta) \rceil$, it follows that $\Pr[Err] \leq \delta$. □

### 4.3.2 Complexity

**Theorem 4.3.5.** *Given an oracle for* SAT*,* ApproxMC$(F, \varepsilon, \delta)$ *runs in time polynomial in* $\log_2(1/\delta), |F|$ *and* $1/\varepsilon$ *relative to the oracle.*

*Proof.* The pseudo-code for ApproxMC can be partitioned into three regions for ease of analysis: (i) lines 1–3, (ii) lines $4 - -9$ (repeat-until loop), and (iii) line 10. Lines 1–3 take time no more than a polynomial in $\log_2(1/\delta)$ and $1/\varepsilon$. The repeat-until loop in lines 4–9 is repeated $t = \lceil 35\log_2(3/\delta) \rceil$ times. The time taken for each iteration is dominated by the time taken by ApproxMCCore. Finally, computing the median in line 10 takes time linear in $t$. The proof is therefore completed by showing that ApproxMCCore takes time polynomial in $|F|$ and $1/\varepsilon$ relative to the SAT oracle.

Referring to the pseudo-code for ApproxMCCore, we find that BoundedSAT is called at most $\mathcal{O}(|F|)$ times. Each such call can be implemented by at most $pivot + 1$ calls to a SAT oracle, and since $pivot + 1$ is in $\mathcal{O}(1/\varepsilon^2)$, the cumulative number of calls to the SAT oracle is a polynomial in $|F|$ and $1/\varepsilon$ relative to the oracle. The random choices in lines 8 and 9 of ApproxMCCore can be implemented in time polynomial in $n$ (hence, in $|F|$) if we have access to a source of random bits. Construction and writing of $F \wedge (h(z_1, \ldots z_n) = \alpha)$ on the memory takes time polynomial in $|F|$. Therefore, the total time taken by each invocation of ApproxMCCore is a polynomial in $|F|$ and $1/\varepsilon$ relative to the SAT oracle. □

## 4.4 Experimental Methodology

To evaluate the performance and quality of results of ApproxMC, we built a prototype implementation and conducted an extensive set of experiments. The suite of benchmarks represent problems from practical domains as well as problems of theoretical interest. In particular, we considered a wide range of model counting benchmarks from different domains including grid networks, plan recognition, DQMR networks, Langford sequences, circuit synthesis, random $k$-CNF and logistics problems [68, 27]. The suite consisted of benchmarks ranging from 32 variables to 229100 variables in CNF representation. The complete set of benchmarks (numbering above 200) is available at http://www.cs.rice.edu/CS/Verification/Projects/ApproxMC/.

Our experiments were conducted on a high-performance computing cluster. Each individual experiment was run on a single node of the cluster; the cluster allowed multiple experiments to run in parallel. Every node in the cluster had two quad-core Intel Xeon processors with 4GB of main memory. We used 2500 seconds as the timeout for each invocation of BoundedSAT in ApproxMCCore, and 20 hours as the timeout for ApproxMC. If an invocation of BoundedSAT in line 10 of the pseudo-code of ApproxMCCore timed out, we repeated the iteration (lines 6-11 of the pseudocode of ApproxMCCore) without incrementing $i$. The parameters $\varepsilon$ (tolerance) and $\delta$ (confidence being $1-\delta$) were set to 0.75 and 0.1 respectively. With these parameters, ApproxMC successfully computed counts for benchmarks with upto 33,000 variables.

We implemented leap-frogging, as described in [24], to estimate initial values of $i$ from which to start iterating the repeat-until loop of lines 6–11 of the pseudocode of ApproxMCCore. To further optimize the running time, we obtained tighter estimates of the iteration count $t$ used in algorithm ApproxMC, compared to those given by algorithm ComputeIterCount. A closer examination of the proof of Theorem 4.3.4 shows that it suffices to have $\eta(t,t/2,0.4) \leq \delta$. We therefore pre-computed a table that gave the smallest $t$ as a function of $\delta$ such that $\eta(t,t/2,0.4) \leq \delta$. This sufficed for all our experiments and gave smaller values of $t$ (e.g., we used $t=41$ for $\delta=0.1$) compared to those given by ComputeIterCount.

For purposes of comparison, we also implemented and conducted experiments with the exact counter Cachet [57] by setting a timeout of 20 hours on the same computing platform. We compared the running time of ApproxMC with that of Cachet for several benchmarks, ranging from benchmarks on which Cachet ran very efficiently to those on which Cachet timed out. We also measured the quality of approximation produced by ApproxMC as follows. For each benchmark on which Cachet did not time out, we obtained the approximate count from ApproxMC with parameters $\varepsilon = 0.75$ and $\delta = 0.1$, and checked if the approximate count was indeed within a factor of 1.75 from the exact count. Since the theoretical guarantees provided by our analysis are conservative, we also measured the relative error of the counts reported by ApproxCount using the $L_1$ norm, for all benchmarks on which Cachet did not time out. For an input formula $F_i$, let $A_{F_i}$ (resp., $C_{F_i}$) be the count returned by ApproxCount (resp., Cachet). We computed the $L_1$ norm of the relative error as $\frac{\Sigma_i |A_{F_i} - C_{F_i}|}{\Sigma_i C_{F_i}}$.

To illustrate the trade-off between confidence ($\delta$) and runtime of ApproxMC, we computed runtime for different values of $\delta$ ($0.4 \leq \delta \leq 0.1$) for a subset of benchmarks. For every benchmark, we normalized the runtime for particular value of $\delta$ by runtime for $\delta = 0.1$. We then average normalized runtime over the subset of benchmarks for every $\delta$ and plot our results.

Since Cachet timed out on the most of the large benchmarks, we compared ApproxMC with state-of-the-art bounding counters as well. As discussed in Section 2.1.2, bounding counters do not provide any tolerance guarantees. Hence their guarantees are significantly weaker than those provided by ApproxMC, and a direct comparison of performance is not meaningful. Therefore, we compared the sizes of the intervals (i.e., difference between upper and lower bounds) obtained from existing state-of-the-art bounding counters with those obtained from ApproxMC. To obtain intervals from ApproxMC, note that Theorem 4.3.4 guarantees that if ApproxMC($F, \varepsilon, \delta$) returns $c$, then $\Pr[\frac{c}{1+\varepsilon} \leq |R_F| \leq (1+\varepsilon) \cdot c] \geq 1-\delta$. Therefore, ApproxMC can be viewed as computing the interval $[\frac{c}{1+\varepsilon}, (1+\varepsilon) \cdot c]$ for the model count, with confidence $\delta$. We also compare the lower and upper bounds of the intervals returned by various bounding counters and ApproxMC. We considered state-of-the-art lower bounding counters, viz. MBound [30], Hybrid-MBound [30], SampleCount [66] and BPCount [27], to compute a lower bound of the model count, and used MiniCount [27] to obtain an upper bound. We observed that SampleCount consistently produced better (i.e. larger) lower bounds than BPCount for our benchmarks. Furthermore, the authors of [30] advocate using Hybrid-MBound instead of MBound. Therefore, the lower bound for each benchmark was obtained by taking the maximum of the bounds reported by Hybrid-MBound and SampleCount.

Our experiments indicated that MiniCount gave incorrect upper bounds when the required confidence was set to values smaller than 0.99. In fact, the authors of MiniCount recommend using a confidence of 0.99, since there are stringent assumptions in the design of the tool. Therefore, we set the confidence bound of MiniCount to the prescribed value of 0.99. The overall confidence of the model count lying in an interval obtained from independently computed lower and upper bounds is the product of the individual confidences. Therefore, to get an overall confidence of 0.9 (for a fair comparison with ApproxMC), we needed SampleCount and Hybrid-MBound to return lower bounds with confidence of $0.9/0.99 \approx 0.91$. The parameter space of SampleCount and Hybrid-MBound allows confidence levels of the form $1 - 2^{\alpha \cdot t}$, where both $\alpha$ and $t$ are positive integers. Therefore, we used $0.875$, the closest lower confidence level (thereby allowing the computed lower bound to be higher), as the confidence of SampleCount and Hybrid-MBound in our experiments. In our comparison of the size of intervals we consider the maximum of lower bounds returned by SampleCount and Hybrid-MBound.

Our implementation of Hybrid-MBound used the "conservative" approach described in [30], since this provides the best lower bounds with the required confidence among all the approaches discussed in [30]. Finally, to ensure fair comparison, we allowed all bounding counters to run for 20 hours on the same computing platform on which ApproxMC was run.

## 4.5 Results

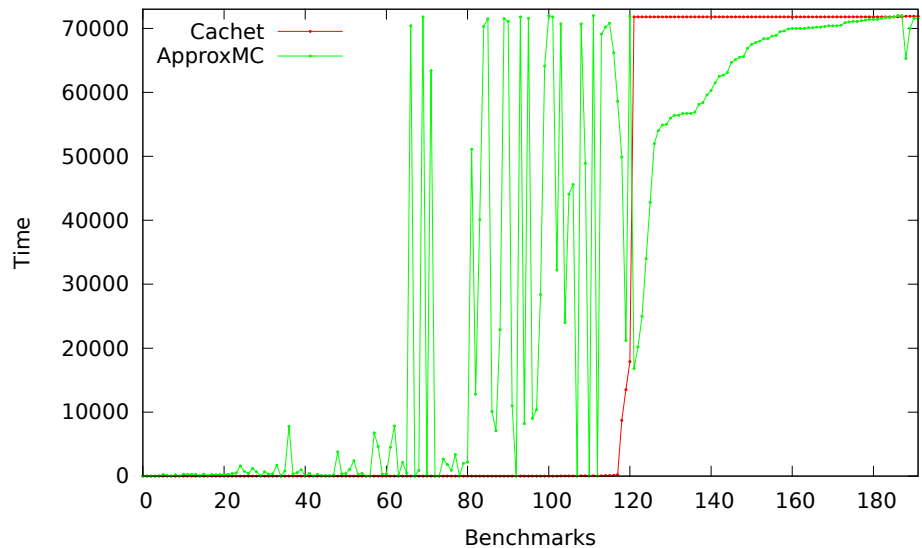

Figure 4.1 : Performance comparison between ApproxMC and Cachet. The benchmarks are arranged in increasing order of running time of Cachet.

Figure 4.1 shows how the running times of ApproxMC and Cachet compared on our set of benchmarks. The y-axis in the figure represents time in seconds, while the x-axis represents benchmarks arranged in ascending order of running time of Cachet. The comparison shows that although Cachet performed better than ApproxMC initially, it timed out as the "difficulty" of problems increased. ApproxMC, however, continued to return bounds with the specified tolerance and confidence, for many more difficult and larger problems. Eventually, however, even ApproxMC timed out for very large problem instances. Figure 4.2 shows the running time of ApproxMC combined with Cachet for timeout of 300 seconds compared with Cachet on the same set of benchmarks. Our experiments clearly demonstrate that there is a large class of practical problems that lie beyond the reach of exact counters, but for which we can still obtain counts with $(\varepsilon, \delta)$-style guarantees in reasonable time. The results suggest that given a model counting problem, it is advisable to run Cachet initially with a small timeout. If Cachet times out, ApproxMC should be run with a larger timeout. Finally, if ApproxMC also times out, counters with much weaker guarantees but shorter running times, such as bounding counters, should be used.

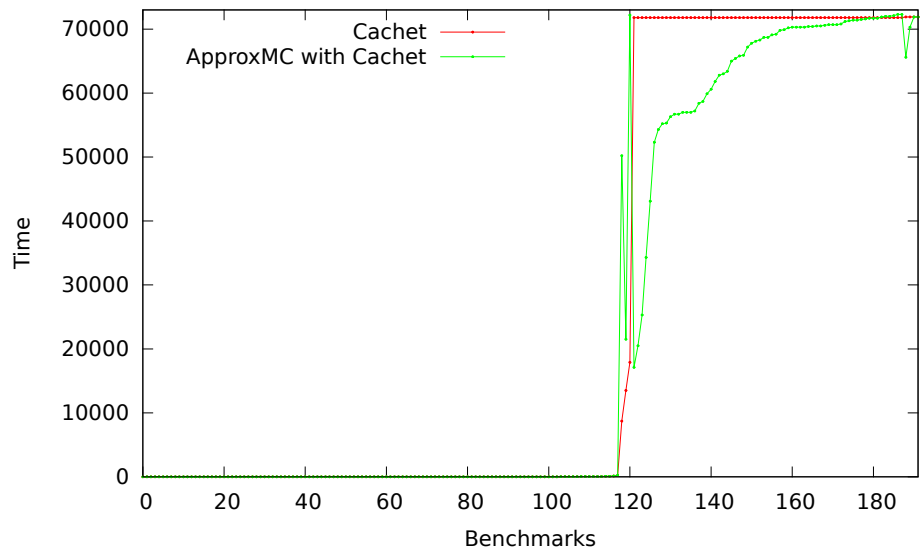

Figure 4.2 : Performance comparison between ApproxMC (with Cachet timeout) and Cachet. The benchmarks are arranged in increasing order of running time of Cachet.

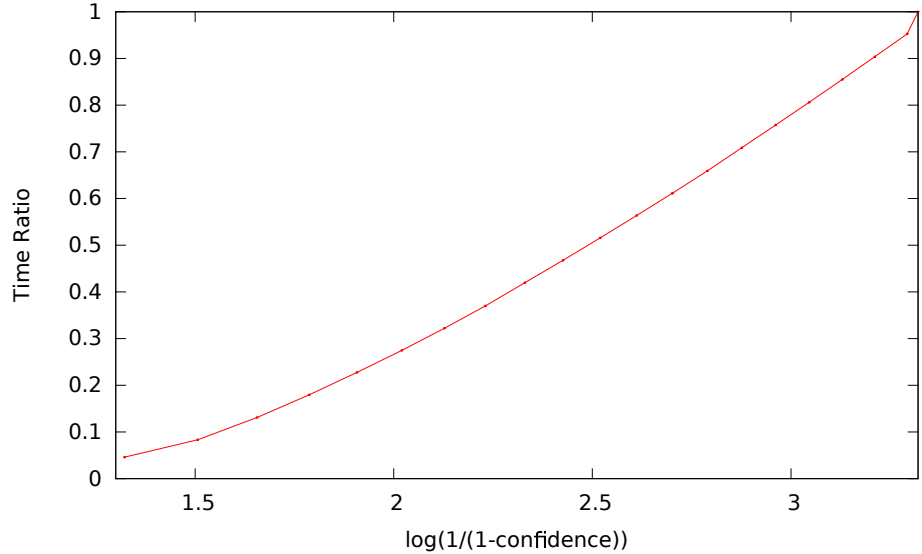

Figure 4.3 : Comparison of normalized runtime with confidence

Figure 4.3 and Figure 4.4 compare the normalized runtime (averaged over a subset of benchmarks) for different values of confidence. The y-axis in both the figures represent normalized runtime (as described in Section 4.4). The x-axis in Figure 4.3 represents $\log(1/\delta)$ while x-axis in Figure 4.4 represents $(1-\delta)$. From both the above two plots, we observe that the runtime increases as the value of confidence increases ($\delta$ decreases, therefore $\log(1/\delta)$ increases), thereby illustrating the relationship suggested by Theorem 4.3.5.

Figure 4.5 shows the observed tolerance (averaged over a subset of benchmarks) for different values of confidence. The y-axis in the figure represents the observed tolerance (as described in Section 4.4), while the x-axis represents confidence $(1-\delta)$.

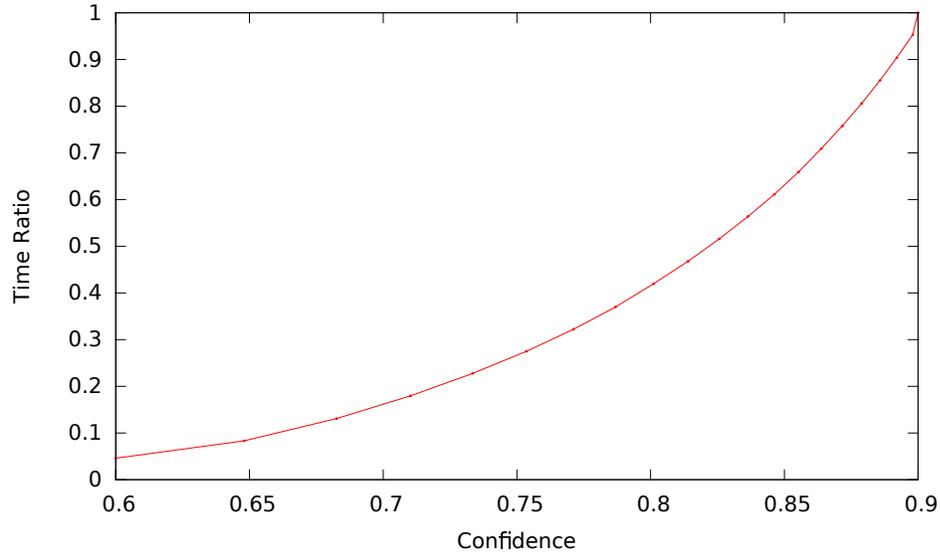

Figure 4.4 : Comparison of normalized runtime with confidence

From the plot, we see that observed tolerance, expectedly, decreases with increase in confidence. It is interesting to note that even for the lowest value of confidence (0.6), the observed tolerance is only 0.16 (lower than allowed tolerance of 0.75), which illustrates the conservative nature of our theoretical analysis.

Figure 4.6 compares the observed tolerance with normalized runtime for a subset of benchmarks. The y-axis in the figure represents the observed tolerance, while the x-axis represents normalized runtime. The plot shows that the observed tolerance decreases as the runtime increases. It is worth noting that the maximum observed tolerance is just 0.16 while our experiments set $\varepsilon$ to 0.75, which indicates that in practice we can obtain counts with desired tolerance from small values of parameter $t$, which determines the number of times ApproxMCCore is invoked.

Figure 4.7 compares the model count computed by ApproxMC with the bounds obtained by scaling the exact count obtained from Cachet by the tolerance factor (1.75) on a subset of our benchmarks for which both Cachet and ApproxMC returned counts (in total). The y-axis in this figure represents the model count on a log-scale, while the x-axis represents the benchmarks arranged in ascending order of the model count. The figure shows that in all cases, the count reported by ApproxMC lies within the specified tolerance of the exact count. Although we have presented results for only a subset of our benchmarks (37 in total) in Figure 4.7 for reasons of clarity, the counts reported by ApproxMC were found to be within the specified tolerance of the exact counts for *all* 95 benchmarks for which Cachet reported exact counts. We also found that the $L_1$ norm of the relative error, considering all 95 benchmarks for which Cachet returned exact counts, was 0.033. Thus, ApproxMC has approximately 4% error in practice – much smaller than the theoretical guarantee of 75% with $\varepsilon = 0.75$.

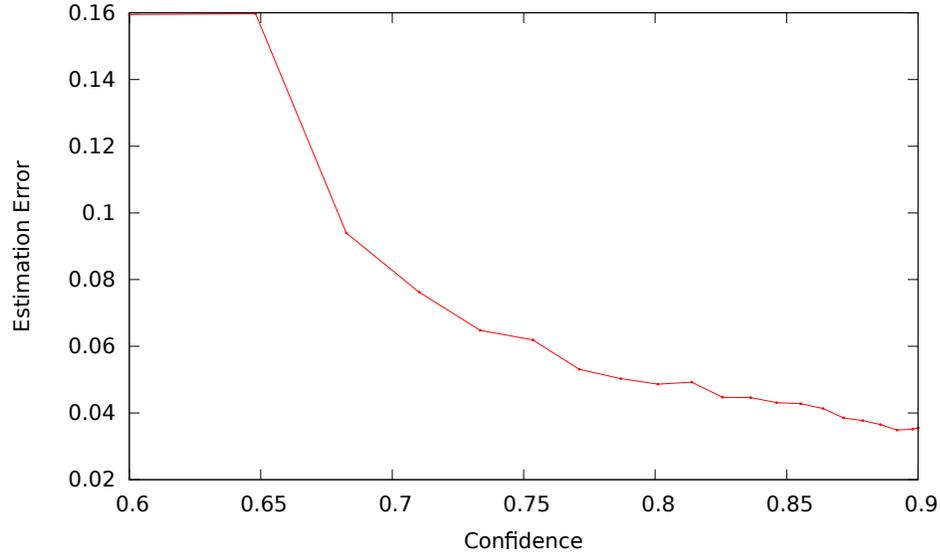

Figure 4.5 : Observed tolerance (averaged over a subset of benchmarks) for different values of confidence

Figure 4.8 compares the sizes of intervals computed using ApproxMC and using state-of-the-art bounding counters (as described in Section 4.4) on a subset of our benchmarks. The comparison clearly shows that the sizes of intervals computed using ApproxMC are consistently smaller than the sizes of the corresponding intervals obtained from existing bounding counters. Since smaller intervals with comparable confidence represent better approximations, we conclude that ApproxMC computes better approximations than a combination of existing bounding counters.

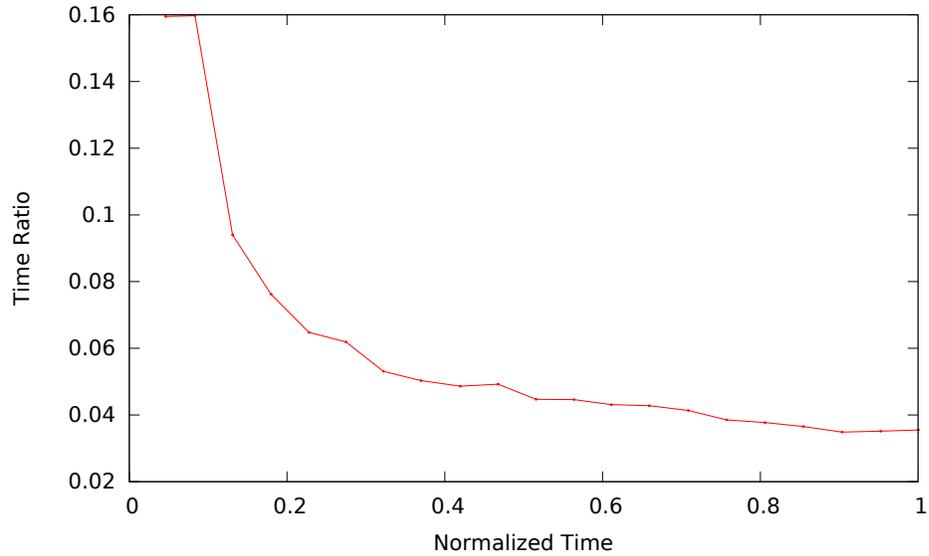

Figure 4.6 : Comparison of observed tolerance with normalized runtime over a subset of benchmarks

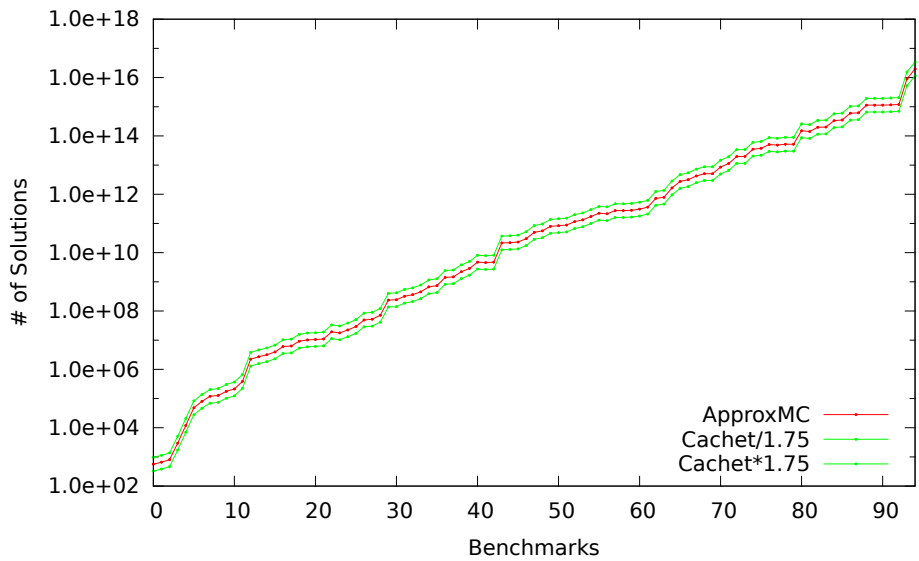

Figure 4.7 : Quality of counts computed by ApproxMC. The benchmarks are arranged in increasing order of model counts.

Figure 4.9 compares the counts returned by ApproxMC and state-of-art bounding counters: SampleCount, MBound, and MiniCount. The comparison clearly shows that in all cases ApproxMC improved the upper bounds from MiniCount significantly; it also improved lower bounds from SampleCount and MBound to a lesser extent. Thus we conclude that not only ApproxMC provides stronger guarantees than the existing bounding counters, it also computes better bounds than the state-or-art bounding counters.

In summary, we showed that it is possible to design an $(\varepsilon, \delta)$ approximate counter for CNF formulae that scales to tens of thousands of variables in practice. To the best of our knowledge, ApproxMC is the first counter in this category.

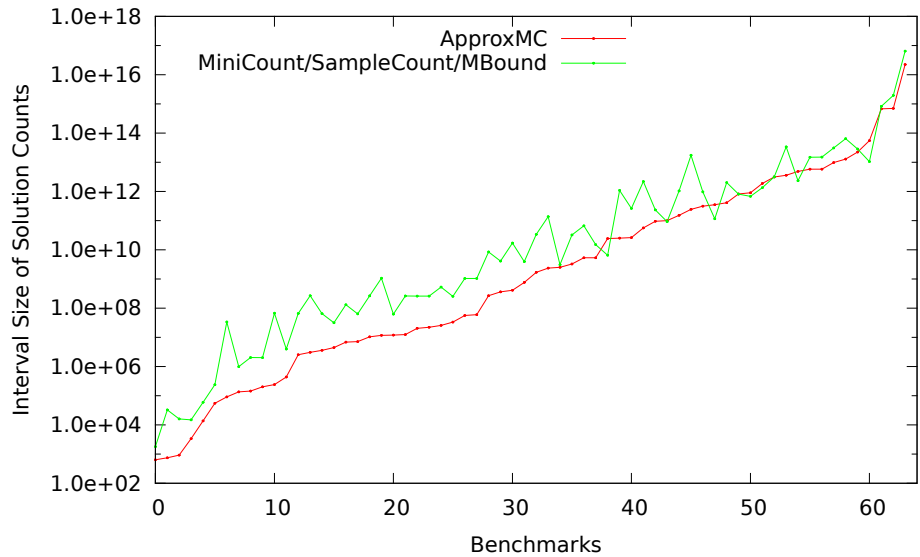

Figure 4.8 : Comparison of interval sizes from ApproxMC and those from bounding counters. The benchmarks are arranged in increasing order of model counts.

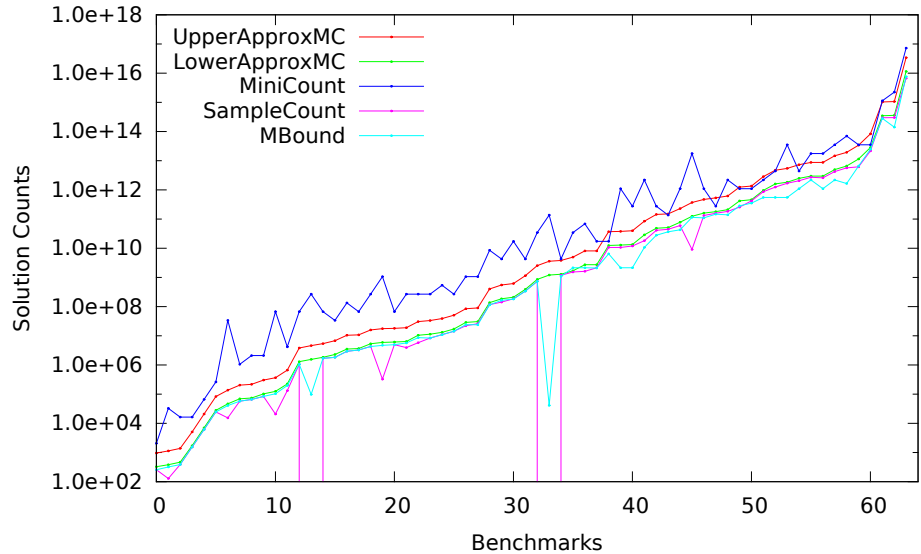

Figure 4.9 : Comparison of counts from ApproxMC and those from bounding counters. The benchmarks are arranged in increasing order of model counts.

## 4.6  Extension to other #P Problems

The techniques presented in this chapter provide approximate model counting for SAT formulae. Since #SAT is #P-complete, therefore, any problem, say Q, in #P can be reduced in polynomial time to a SAT formula, say F, such that number of models of Q is equal to number of models of F. Therefore, the algorithms presented in this chapter can solve any problem in #P.

# Chapter 5

# Conclusions and Future Work

This thesis proposes techniques for two related problems of significant theoretical and practical interest: near uniform generation of witnesses of CNF SAT formulas, and approximately counting models (or witnesses) of these formulas. In this chapter, we summarize the main contributions of this thesis and outline directions for future research.

## 5.1  Summary of Contributions

Although both (near) uniform generation and approximate model counting have received attention in the past, prior work either offered very weak or no guarantees or did not scale to real world examples. The primary contribution of this thesis is a novel approach based on limited-independence hashing that allows us to design algorithms for both problems, with strong theoretical guarantees and scalability extending to formulas with thousands of variables. The proposed approach differs from existing hashing based techniques in three key ways: (i) the use of computationally efficient linear hashing functions with low degrees of independence as opposed to computationally inefficient algebraic hash functions with higher degrees of independence (ii) the requirement of only a randomly chosen hash cell to be "small" instead of requiring *every* cell to be "small", and (iii) reduction in the size of "small" cells from $n^2$ in the previous work [18] to $n^{1/k}$ for $2 \leq k \leq 3$, where n is the number of variables. These dif-ferences allow our algorithms to scale to problems several orders of magnitude larger than what was earlier possible, while still ensuring strong theoretical guarantees of uniformity.

In the first part of the thesis, an algorithm named UniWit was presented for generating witnesses of CNF SAT formulas near uniformly in the space of all satisfying assignments. Near-uniformity of the generated witnesses was proved theoretically, while the practical utility of the algorithm was empirically demonstrated through extensive experiments. Our experimental results indicate that UniWit outperforms existing state-of-the-art algorithms, both in terms of runtime and uniformity of the generated witnesses.

In the second part of thesis, a scalable approximate model counting technique for CNF formulas called ApproxMC was presented. To the best of our knowledge, ApproxMC is the first scalable approximate model counter for CNF formulas. Given a tolerance in $(0,1)$ and a confidence measure, our theoretical analysis shows that ApproxMC provably achieves the specified tolerance in reporting the model count with the specified confidence. Experiments on a large suite of benchmarks from diverse domains show that ApproxMC scales to problems involving tens of thousands of variables.

## 5.2  Looking ahead

In light of the findings of this thesis, there are several interesting questions that remain to be answered. We outline some of them below, indicating directions for future research.

**Stronger guarantees**

As discussed in Chapter 1, Jerrum et al. have shown that for CNF SAT, the problem of generating satisfying assignments almost uniformly is polynomially inter-reducible with approximate model counting [14]. Significantly, their proof shows that an almost-uniform generator can be used as a black-box in the design of an algorithm for approximate model counting and vice versa. Our limited-independence hashing based approach gives an algorithm for approximate model counting and also an algorithm for near-uniform generation. While near-uniform generation is a more relaxed notion than almost uniform generation, our algorithm makes crucial uses of key steps in the near-uniform generation algorithm; hence it uses the near-uniform generator as a white-box, instead of a black-box. This leaves open the question of whether a near-uniform generator can be used as a black-box to design an $(\varepsilon, \delta)$ approximate model counter and, hence (using Jerrum etal's result) an almost uniform generator. Another related direction of future research is to investigate if an approximate counter can be used to design a scalable generator with strong guarantees of uniformity.

**Scaling to larger problems**

This thesis made the first steps towards designing scalable near-uniform generators and approximate model counters capable of handling tens of thousands of variables. Many real-world applications, however, continue to be out of the reach of proposed algorithms. Hence, extending our approach to handle very large problems is an important challenge that needs to be addressed. It is well known that adding random XOR constraints to a CNF formula makes it harder to find satisfying assignments to the formula [69]. Previous work in the area of linear hash functions [69] suggests that by restricting each random XOR clause to refer to only a few variables, the difficulty of finding SAT assignments of the overall constraints can be mitigated in practice. However, such restrictions fail to provide independence guarantees. Therefore, the design of efficiently computable linear hash functions that refer to only a few variables in each clause and provide required independence guarantees is an important problem. Specifically, a solution to this problem would translate to improved scalability of our algorithms. Promising steps have been taken in recent works [70, 71] towards the design of efficient hash functions.

**Richer constraints**

Richer constraint languages (e.g., SMTLIB) provide the ability to specify constraints arising from real-world problems in a user friendly and succinct form. A wide variety of industrial applications are encoded in such rich constraint languages. Our current framework requires translation of constraints to Boolean constraints. An interesting and practically useful extension of the current work would be to consider richer constraint languages and to build approximate counters and (near)-uniform generators modulo theories, leveraging recent progress in satisfiability modulo theories, c.f., [72]. Such an endeavor would also require designing efficient limited-independence hash functions and appropriate constraint solvers for such richer constraint languages.

**Extension to other problems in NP and #P**

While the algorithms presented in this thesis focused on SAT formulas, the core ideas are quite general and can be extended to other problems in $\mathcal{NP}$ and #P. One possible direction, as outlined above, is through designing appropriate hash functions and constraint solvers. Another promising direction is to utilize $\mathcal{NP}$-completeness of SAT and #P-completeness of #P. While, as discussed in Section 4.6, the parsimonious reduction for #P allows us to employ our technique to perform model counting for any #P problem but a reduction for $\mathcal{NP}$ is not necessarily parsimonious. Therefore known polynomial time reduction for a $\mathcal{NP}$ problem to SAT may not preserve the number of solutions, thus rendering us unable to use our current algorithms. Nevertheless models of the original problem do agree with those of the transformed SAT problem projected on a subset of variables. Therefore, an interesting direction of research would be to explore if we can perform uniform sampling for assignments projected on a subset of variables than all the variables [71].

**Weighted uniform generation and approximate counting**

Several applications such as probabilistic inference can be reduced to weighted model counting [22]. Similarly, simulation based techniques benefit from the ability to handle user provided bias, which can be reduced to weighted uniform generation. Thus, extension of the current approach to weighted generation and counting is an important and interesting direction for future research.